# The role of dendritic spines in water exchange measurements with diffusion MRI: Time-Dependent Single Diffusion Encoding MRI

Kadir Şimşek[1,2] Arthur Chakwizira[3], Markus Nilsson[4] and Marco Palombo[1,2]

[1]Cardiff University Brain Research Imaging Centre (CUBRIC), Cardiff University, Cardiff, United Kingdom

[2]School of Computer Science and Informatics, Cardiff University, Cardiff, United Kingdom

[3]Medical Radiation Physics, Lund, Lund University, Lund, Sweden

[4]Department of Clinical Sciences Lund, Radiology, Lund University, Lund, Sweden

**Corresponding Author**: Kadir Şimşek

Cardiff University Brain Research Imaging Centre (CUBRIC), School of Psychology, Cardiff 'University, Cardiff, United Kingdom,

Maindy Road, CF24 4HQ, Cardiff, United Kingdom

+44 2920874000 ext:20063 | simsekk@cardiff.ac.uk

**Word count: 8946**

**Sponsors/Grant numbers:**

- UKRI Future Leaders Fellowship
  - (MR/T020296/2)
- Swedish Research Council (2024-04968)
- Hjärnfonden (FO2024-0335-HK-73)

# Abstract

Time-dependent diffusion MRI (dMRI) using single diffusion encoding (SDE) is sensitive to water dynamics in biological tissues, yet interpreting its signals requires careful consideration of underlying microstructure. While prior work has focused on restricted/hindered diffusion and membrane permeation, additional exchange mechanisms such as diffusion-mediated exchange between dendritic shaft and spines in gray matter (GM) remain understudied.

Here, we hypothesize that water diffusion within impermeable spiny dendrites can produce time-dependent SDE signals indistinguishable from those arising from permeative exchange; and assess to what extent spine density impacts estimates of exchange time.

Using Monte Carlo simulations and analytical solutions from the narrow escape problem, we quantify spine–shaft and shaft–spine exchange times, revealing characteristic times (3–26 ms) comparable to permeative exchange estimates in the cortex. We show that a modified two-compartment Kärger model accurately captures the time-dependent SDE signal along spiny dendrites but yields biased exchange estimates, that reflect total spine volume fraction rather than specific spine morphology. Simulations reveal that unaccounted diffusion-mediated exchange could introduce up to 80% bias in Neurite Exchange Imaging (NEXI) and Standard Model with Exchange (SMEX) estimates. Further, we propose an extended three-compartment Kärger model incorporating both diffusion-mediated exchange between dendritic shaft and spines and permeative exchange with extracellular space. Critically, while the three-compartment Kärger model captures both exchange mechanisms, it cannot uniquely disentangle membrane permeability from spine volume fraction.

These findings underscore the necessity of considering dendritic spine contributions when interpreting time-dependent SDE data and caution against attributing exchange effects solely to membrane permeability. Our study further highlights the need for advanced acquisition and modeling approaches to differentiate permeative and diffusion-mediated exchange in GM.

# List of Abbreviations

$\delta$: diffusion gradient duration

$\Delta$: diffusion gradient separation time

$\Delta\tau_{sp\rightarrow sh}$ : percentage change of spine-to-shaft residence time

$\varepsilon$: radial distance to the shaft boundary

$\eta$: concentration curve of spins

$\theta$: angle between diffusion gradient and dendritic shaft

$\lambda_{und}$: undulation wavelength

$\nu$: total spine volume fraction

$\xi$: membrane permeability

$\rho$: spine density along the dendritic shaft

$\sigma$: pore area density

$\tau$: residence time

$\tau_{e\rightarrow i}$: residence time from extra-to-intracellular space

$\tau_{i\rightarrow e}$: residence time from intra-to-extracellular space

$\tau_{sh\rightarrow sp}$: shaft-to-spine residence time

$\tau_{sp\rightarrow sh}$: spine-to-shaft residence time

$A_{bead}$: beading amplitude

$ADC$: apparent diffusion coefficient

CA1: hippocampal cornu ammonis 1

$D$: diffusivity

$D_e$: extracellular diffusivity

$D_i$: intracellular diffusivity

$D_{in}$: intra-neurite diffusivity

$D_{sp}$: diffusivity in dendritic spines

DDE: double diffusion encoding

DEXSY: diffusion exchange spectroscopy

DM: diffusion-mediated exchange

dMRI: diffusion-weighted MR imaging

EM: electron microscopy

$f$: intracellular volume fraction

FWF: free gradient waveform

$\vec{g}$: diffusion gradient vector

GM: gray matter

$k$: exchange rate

$K$: kurtosis

$k_{ex}^{DM}$: diffusion-mediated exchange

$k_{sh \to sp}$: shaft to spine exchange rate

$k_{sp \to sh}$: spine to shaft exchange rate

$l_c$: correlation length

$L_{neck}$: spine neck length

$L_{shaft}$: dendritic shaft length

$\vec{n}$: neurite orientation

$N_s$: number of spins

$N_t$: number of time steps

$N_{und}$: undulation period along the shaft

NET: narrow escape problem

NEXI: Neurite Exchange Imaging

np: narrow pulse

PGSE: pulsed gradient spin echo

$R_c$: radius of curvature at the spine head and neck connection

$R_{head}$: spine head radius

$R_{neck}$: spine neck radius

$R_{shaft}$: dendritic shaft radius

$S_{||}$: diffusion signal parallel to the dendritic shaft

$S_{\perp}$: diffusion signal perpendicular to the dendritic shaft

$S_e$: diffusion signal in the extracellular space

$S_{sh}$: diffusion signal in the dendritic shaft

$S_{sp}$: diffusion signal in the dendritic spine

SDE: single diffusion encoding

SMEX: Standard Model with Exchange

SVR: Surface to volume ratio

$t_{ex}^{NEXI}$: extimated exchange time with NEXI

$t_d$: diffusion time

$t_{ex}$: exchange time

$t_{ex}^{DM}$: diffusion-mediated exchange time

$V_{head}$: Spine head volume

WM: white matter

wp: wide pulse

# 1 Introduction

Time-dependent diffusion MRI (dMRI) is a powerful technique that probes the temporal dynamics of molecular diffusion within biological tissues, offering insights into both restriction and exchange[1–15]. Numerous methods have been proposed to quantify exchange rates of water molecules in the brain[16–30]. Some studies investigated exchange rates using model-based approaches, under the assumption that permeative exchange across the cellular membrane is the most relevant exchange process. While permeative exchange in white matter (WM) is expected to be slow and negligible due to the insulating myelin sheath surrounding axons[14,31–33], it is likely faster and non-negligible in gray matter (GM), where myelinated neurites (axons and dendrites) are far less abundant. Two recent examples of such model-based methods, designed to characterize permeative exchange in GM, are the Neurite Exchange Imaging (NEXI)[14] and the Standard Model with Exchange (SMEX)[13]. Both NEXI and SMEX rely on the pulsed gradient spin echo (PGSE) acquisitions[34] with multiple diffusion times, i.e. single diffusion encoding (SDE) measurements with different diffusion times. Using NEXI/SMEX, recent works characterized water exchange in GM, reporting exchange times <100 ms [13,14,35,36], which are comparable to the typical diffusion times used in SDE measurements.

However, the SDE signal at varying diffusion times carries entangled information on both restriction and exchange processes[2,13,14,35,36]; therefore, disentangling and estimating diffusion metrics specific to either restriction or exchange remains degenerate[6,25,37–39]. Recent studies exploited advanced diffusion encoding techniques to address this degeneracy in time-dependent SDE measurements. One approach involves double diffusion encoding (DDE) measurements which help to disentangle this degeneracy by introducing additional experimental dimensions[2,18,28,40,41]. Other studies use diffusion exchange spectroscopy (DEXSY) methods to isolate distinct water pools with different restriction and exchange characteristics and quantify the corresponding characteristic restriction size and exchange times[27,28,30]. Others modulated the DDE blocks to isolate different exchange rates and characterize them[17,18,22]. In addition to DDE methods, more complex encoding schemes like free gradient waveforms (FWF) have been utilized to disentangle restriction and exchange in the brain, particularly in the GM[24,25,42]. One recent study mapped how sensitive the diffusion signal acquired with FWF encoding is to restriction and exchange effects and provided sequence optimization for determining the interplay between restriction and exchange effects[24].

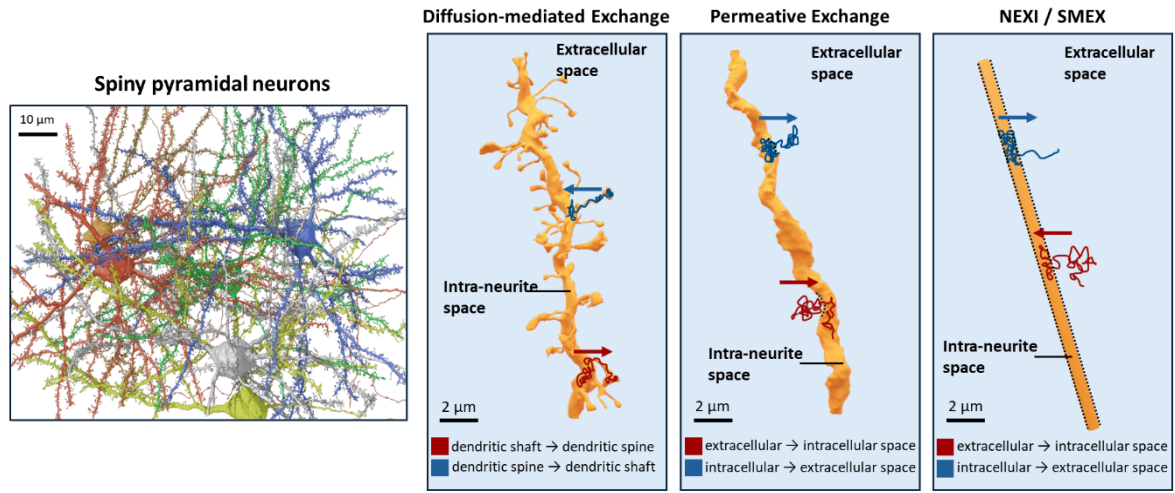

**Figure 1**: *Illustration of spiny dendritic branches and different diffusion exchange mechanisms. The diffusion exchange mechanism is often considered as permeative exchange across cell membrane between intracellular and extracellular compartments. The NEXI and SMEX are the well-known biophysical models to estimate exchange times between these compartments. Here we introduce another mechanism contributing to the exchange estimates, referred to as diffusion-mediated exchange which is the exchange between intracellular sub-compartments (e.g. between dendritic spines and dendritic shaft).*

While advanced diffusion encodings (beyond SDE) show promise for disentangling restriction and exchange, time-dependent SDE measurements remain clinically appealing due to their experimental robustness[43]. However, interpreting these measurements requires careful consideration of the underlying microstructure. Although time-dependent SDE signals decreasing with time at a fixed b-value are often attributed to permeative exchange, growing evidence suggests contributions from other exchange mechanisms. Some recent studies reported on potential alternative exchange mechanisms besides permeability, e.g. geometric exchange within impermeable compartments, and vascular water exchange[13,16,17,23,24,28,44,45] and some preliminary works suggest that dendritic spines (small, protruding structures on the dendrites of neurons), due to their restrictive geometry, may exhibit time-dependent signatures resembling permeative exchange[46–48], highlighting the need to distinguish these mechanisms in SDE-based analyses.

The aim of this work is to investigate the degree to which subcellular structures such as dendritic spines contribute to estimates of exchange in GM from time-dependent SDE measurements. We hypothesize that water diffusing within spiny dendrites in the GM, without crossing the cell membrane, can impart a time-dependent signature which is similar to or possibly indistinguishable from permeative exchange (see **Figure 1**). Moreover, we propose an extended Kärger model of three compartments: dendritic shaft, spine and extracellular space, inspired by a previous model for nerves[49]. The signal from spins diffusing within the shaft is modeled as diffusion within sticks. The signal within the spines is treated as fully restricted, representing still water, while the extracellular space is modeled as isotropic Gaussian diffusion. All compartments interact and exchange with each other. Using Monte-Carlo simulations and digital substrates of spiny neurites, we test our hypothesis, evaluate the impact of different spine densities and morphologies on the diffusion exchange mechanisms and compare it with permeative exchange estimates using NEXI and SMEX.

## 2 Theory

In this section, our objective is to develop a theoretical framework characterizing both permeative and diffusion-mediated exchange processes. By deriving analytical predictions, we will establish a benchmark for interpreting our results from Monte Carlo simulations (described in the Methods section). This approach will enable us to elucidate the interplay between these two exchange mechanisms in time-dependent SDE measurements, providing deeper insights into their relative contributions and dynamics.

We first provide an overview of the dendritic spines' morphology from existing literature and then introduce theoretical approaches to characterize both molecular diffusion in smooth dendrites subject to permeative exchange across the dendritic membrane, and molecular diffusion in spiny dendrites, accounting for diffusion-mediated exchange between dendritic shaft and spines.

### 2.1 Dendritic spines: morphological diversity and regional variations

Dendritic spines are small, protruding structures on the dendrites of neurons which were first revealed by Cajal using Golgi staining technique[50]. They exhibit diverse morphologies, including thin, stubby, and mushroom-shaped forms[51–53], which are associated with synaptic strength and plasticity[54–57]. Their shape and density can vary in response to neural activity, influencing cognitive functions and neurological disorders across the whole brain[56,58–60]. On average, a dendritic spine has a volume ranging from 0.01 to 0.8 $\mu m^3$, with surface area varying between 0.1 to 5 $\mu m^2$ [58,61,62]. The spine head radius, often associated with synaptic strength, typically ranges from 0.15 to 0.55 $\mu m$, with mushroom spines having the largest spine heads[61,63,64]. The spine neck has a length ranging from 0.1 to 2 $\mu m$ and a radius ranging from 0.02 to 0.25 $\mu m$ [53,62,63]. The dendritic spine sizes offer structural basis for understanding synaptic integration and plasticity.

The spine density on the mature neurons ranges between 1 to 10 spines per micron ($\mu m^{-1}$)[58,65–67]. For instance, hippocampal thick apical dendrites of CA1 pyramidal neurons near the stratum radiatum can exhibit spine density up to 10 $\mu m^{-1}$ [58,68–71] while the pyramidal neurons in the visual cortex have spine densities around 1 $\mu m^{-1}$ or less[72,73]. The pyramidal neurons in the cortical layers can exhibit different spine densities due to their distinct dynamics[74–77]. The spiniest type of neurons are the cerebellar Purkinje cells. 3D electron-microscopy (EM) analysis revealed that the spine density can reach up to 15 μm⁻¹ in the dendritic branches of Purkinje cells[58,78]. In the brain cortex, the estimated synapse density is approximately $7.6x10^8$ synapses per mm³ [79,80]. Assuming each synapse is associated with a dendritic spine of mean volume 0.1 μm³ [53,81–83], we would expect a total volume fraction occupied by spines ($\nu$) in the brain cerebral cortex of $7.6x10^8 *$ $0.1\ x10^{-9} \sim 8\%$. This aligns with reported total volume fractions of dendritic spines in GM of brain cortex over different species[84]. A similar calculation could be done for the cerebellar cortex where the spiniest Purkinje cells form complex synaptic wiring; however, the literature reporting on the synaptic density in the cerebellar cortex is discordant and here we can only suggest that the total volume fraction occupied by spines in cerebellar cortex should be higher than ~8%.

## 2.2   The narrow escape problem and diffusion-mediated exchange in dendritic spines

The molecular diffusion in membrane-enclosed cellular domains with narrow openings has been extensively investigated in the literature[85,86]. The mathematical problem involving diffusing particles escaping through a narrow channel is referred to as the 'Narrow Escape Problem'[87–92]. The problem characterizes the narrow escape time (NET), of a particle inside a composite domain with a narrow pore[86,93–95]. For our problem of characterizing the diffusion of molecules between dendritic shaft and dendritic spines, the narrow openings are the circular openings connecting the shaft to the spine head via the spine neck.

The NET of particles diffusing in membrane-enclosed domains and the related characteristic escape times ($\tau$) have been widely studied[93–95]. Holcman et al. proposed a model to characterize the NET of molecules diffusing from inside the head of a dendritic spine to the shaft connected via the spine's neck, also considering the impact of smooth and non-smooth connections between the dendritic spine's head with its neck[93]. In our work, for simplicity, we used the asymptotic solution for the escape time from spine-to-shaft, $\tau_{sp \to sh}$, of diffusing particles within a spherical dendritic spine connected with its neck at a right angle (i.e. non-smoothly), given by the following equation[93]:

$$\tau_{sp \to sh} = \frac{V_{head}}{4R_{neck}D}\left[1 + \frac{R_{neck}}{\pi R_{head}}\log\left(\frac{R_{head}}{R_{neck}}\right)\right] + \frac{\mathcal{O}(1)}{D} + \frac{L_{neck}^2}{2D} + \frac{V_{head}L_{neck}}{\pi R_{neck}^2 D} \qquad (1)$$

For a general spine morphology connected smoothly to the neck,

$$\tau_{sp \to sh} = \frac{1}{\sqrt{2}}\left(\frac{2R_c}{R_{neck}}\right)^{\frac{3}{2}}\frac{V_{head}}{R_c D}\left(1 + \mathcal{O}(1)\right) + \frac{L_{neck}^2}{2D} + \frac{V_{head}L_{neck}}{\pi R_{neck}^2 D} \qquad (2)$$

where $V_{head}$ is the spine head volume with corresponding radius $R_{head}$. The cylindrical neck has a radius $R_{neck}$ and length $L_{neck}$. $R_c$ is the radius of curvature at the spine head and neck connection. $D$ is the molecules' diffusion coefficient. The $\mathcal{O}(1)$ term can be computed from the explicit Neuman-Green function for spheres[94]. In our work, we used the formulation in Eq.(1) (neglecting the $\mathcal{O}(1)$ term as much smaller than the other terms) to compute the escape time $\tau_{sp \to sh}$ analytically for our digital substrate modelling dendritic spines, and compared the obtained values with those obtained from the Monte-Carlo simulations (see Methods section for more details).

The residence time of a particle inside dendritic spines, as described in Eq.(1), depicts one half of the problem. The other side of the medallion is determining the NET for the particles diffusing from the dendritic shaft to the spine, which is equally important, as the rate of exchange observed by diffusion accounts for the cumulative rate yielded by these two processes. The problem can be considered as an aspect of the narrow escape problem and has been studied analytically for various geometric domains[87–89,92]. One recent work of relevance for this study solved the problem for cylindrical domains[96]. The proposed solution (Eq.20 in ref.[96]) depends on the pore area density ($\sigma$) and the radial distance from the cylindrical boundary ($\varepsilon$), given the condition $\varepsilon = \frac{\pi R_{neck}}{4}$ (where $R_{neck}$ is the spine neck radius and $R_{shaft}$ is the dendritic shaft radius; given that $R_{neck} \ll R_{shaft}$) to satisfy the approximation in Berg and Purcell's original model for the narrow escape problem[92]. The analytical solution for the NET from shaft-to-spine, $\tau_{sh \to sp}$, can then be described as (Eq.20 in ref.[96]):

$$\tau_{sh \to sp} = \frac{R_{shaft}^2}{4D} + \frac{(1-\sigma)(2\varepsilon R_{shaft} - \varepsilon^2)}{4D\sigma} \qquad (3)$$

The pore area density $\sigma$ can be easily computed from the ratio between the total spine neck area on the dendritic shaft and the surface area of dendritic shaft, yielding the following relation $\sigma = \frac{\rho R_{neck}^2}{2R_{shaft}}$ where $\rho$ is the dendritic spine density (i.e., number of spines per unit length of dendritic segment) on the dendrite's branch.

The Eq.(3) completes the whole picture in determining the diffusion-mediated exchange times between intracellular sub-compartments. The diffusion-mediated exchange time between intracellular sub-compartments can be computed using two analytical solutions in Eq.(1) (or Eq.(2)) and Eq.(3). The total exchange rate of the system, $k_{ex}^{DM}$, will be defined by the sum of the exchange rates of each process, $k_{ex}^{DM} = k_{sp \to sh} + k_{sh \to sp}$ where $k = 1/\tau$. Then, a characteristic time for the diffusion-mediated exchange of the system, $t_{ex}^{DM}$, can be computed as $t_{ex}^{DM} = \frac{1}{k_{ex}^{DM}}$. We computed $t_{ex}^{DM}$ to compare it with the characteristic time for the permeable exchange of the system as defined by the Karger model and assess to what extent diffusion-mediated exchange can bias estimates of permeable exchange (see Methods section for more details).

## 2.3 The Kärger model to characterize permeative exchange

The quantification of exchange rates between different water compartments in the brain is essential for biophysical modeling of the diffusion MRI signal. When water compartments are assumed to be separated by permeable membranes, the exchange between these compartments is referred to as permeative exchange. Kärger proposed a model accounting for the exchange between two water pools[1,97]. The model comprises two Gaussian diffusion pools with volume fractions $f_1$ and $f_2$, and includes intercompartmental exchange rates $k_1$ and $k_2$. The exchange rates satisfy the condition $f_1 k_1 = f_2 k_2$, ensuring equilibrium of flux between compartments. Typically, for application to time-dependent SDE measurements in biological tissues, the two compartments are assumed to be the intra and extracellular compartments, from now on defined by the subscripts 'i' and 'e', respectively. The Kärger model assumes that the exchange during the diffusion encoding time is negligible, meaning that the exchange time ($t_{ex} = \frac{1}{k_{i \rightarrow e} + k_{e \rightarrow i}}$) is significantly longer than the duration of the diffusion encoding gradient ($\delta$). Another assumption is that the intracellular compartment is homogenous at all times meaning that the exchange is 'barrier limited', (i.e., $l_c^2 \ll \frac{D_i}{k_{i \rightarrow e}}$), where $l_c$ is the correlation length of the intracellular compartment, $D_i$ is the diffusivity within the compartment, and $k_{i \rightarrow e}$ is the forward exchange rate of the corresponding compartment[2,12,98]. In other words, the diffusing particles fully explore their local compartment before transitioning to another.

Biophysical models accounting for the exchange times in GM that were proposed previously, such as NEXI and SMEX, rely on the Kärger model[13,14]. They consist of two exchanging Gaussian compartments; (i) isotropically oriented neurite (stick) compartment and (ii) the extracellular compartment. The extracellular compartment is assumed to be isotropic due to overall negligible fractional anisotropy in GM[14], which reduces the dimensionality of the model ($D_e \equiv D_{e\perp} = D_{e\parallel}$). The diffusion signal can be computed as the solution of the coupled differential equations for each compartment:

$$\frac{d}{dt} \begin{bmatrix} S_i(t) \\ S_e(t) \end{bmatrix} = \left( \begin{bmatrix} -1/\tau_{i \rightarrow e} & 1/\tau_{e \rightarrow i} \\ 1/\tau_{i \rightarrow e} & -1/\tau_{e \rightarrow i} \end{bmatrix} - q^2(t) \begin{bmatrix} D_i \eta^2 & 0 \\ 0 & D_e \end{bmatrix} \right) \begin{bmatrix} S_i(t) \\ S_e(t) \end{bmatrix} \qquad (4)$$

where $\eta = \hat{g} \cdot \hat{n}$ and $\mathrm{q}(t) = \gamma \int_0^t g(t')dt'$, with $\boldsymbol{g}(t) = \hat{g}g(t)$ being the diffusion-sensitizing gradient, $\gamma$ the proton gyromagnetic ratio, $\hat{n}$ the main neurite's direction, and $\tau_{i \to e}$ and $\tau_{e \to i}$ representing the exchange times respectively from intracellular to extracellular compartments and vice versa. Applying the barrier limited exchange condition $\frac{f}{\tau_{i \to e}} = \frac{1-f}{\tau_{e \to i}}$ for the two Gaussian pools (e.g., neurite and extracellular compartments), one can solve Eq.(4) to obtain the total signal $\mathcal{K}(q, t, \hat{g} \cdot \hat{n}; f, D_i, D_e, t_{ex})$ for SDE acquisition, and use it to compute numerically the direction-averaged SDE signal[14]:

$$\tilde{S}(q,t) = S_0 \int_0^1 \mathcal{K}(q, t, \theta; f, D_i, D_e, t_{ex}) d(\cos \theta) \qquad (5)$$

$$\mathcal{K}(q, t, \hat{g} \cdot \hat{n}; f, D_i, D_e, t_{ex}) = f' e^{-D_i' q^2 t} + (1 - f') e^{-D_e' q^2 t} \qquad (5a)$$

$$D_{i,e}' = \frac{1}{2} \left[ D_i + D_e + \frac{1}{q^2 t_{ex}} \mp \sqrt{\left( \left( D_e - D_i + \frac{2f-1}{q^2 t_{ex}} \right)^2 + \frac{4f - 4f^2}{q^4 t_{ex}^2} \right)} \right] \qquad (5b)$$

$$f' = \frac{f D_i + (1-f) D_e - D_e'}{D_i' - D_e'} \qquad (5c)$$

where $\theta$ is the angle between $\hat{g}$ and $\hat{n}$; $t$ is the diffusion time ($\Delta - \frac{\delta}{3}$, where $\Delta$ is the diffusion gradient separation time and $\delta$ is its gradient duration for the SDE), and $q$ is defined from the b-value as $q = \sqrt{\frac{b}{t}}$.

## 2.4 Modified two-compartments Kärger model to characterize diffusion-mediated exchange between dendritic shaft and spines

In this study, we aim to investigate how well the modified Kärger model can characterize the time-dependent SDE intra-cellular signal decay for a system of impermeable spiny dendrites. We also aim to investigate how accurate the estimates of residence times are, in the absence of a better theory to characterize diffusion-mediated exchange. Assuming radii of dendrites below the resolution limit[99], such that the signal decay perpendicular to the dendritic axis is negligible ($S_\perp \approx 1$), we hypothesize that the modified Kärger model introduced by refs.[12,100] would effectively characterize the impact of diffusion-mediated exchange between dendritic shaft and spines on the intra-cellular diffusion-weighted signal parallel to the dendritic axis, $S_\parallel$. However, we note that the Kärger model is technically incapable of modelling the exchange in spines scenario, because the transition rate between spines and shaft is very high, meaning that the system is likely not in the barrier-limited regime. In such a scenario, the concentration of spins that have never left the compartment falls below the equilibrium concentration, leading to a transport rate that is below the equilibrium one (and depends on time). Hence, we expect discrepancy between the exchange times from our simulation experiments and those predicted from the modified Kärger model.

Worth noting, here we are focusing our analysis on intra-cellular diffusion only. The proposed modified two-compartments Kärger model is made of one Gaussian diffusion compartment (i.e. diffusion along the shaft main axis) and one compartment of still water or dot compartment (i.e., water fully restricted inside the spines, $D_{sp} = 0 \frac{\mu m^2}{ms}$):

$$\frac{d}{dt}\begin{bmatrix} S_{sh,\parallel}(t) \\ S_{sp,\parallel}(t) \end{bmatrix} = \left( \begin{bmatrix} -1/\tau_{sh\rightarrow sp} & 1/\tau_{sp\rightarrow sh} \\ 1/\tau_{sh\rightarrow sp} & -1/\tau_{sp\rightarrow sh} \end{bmatrix} - q^2(t) \begin{bmatrix} D_{sh,\parallel} & 0 \\ 0 & 0 \end{bmatrix} \right) \begin{bmatrix} S_{sh,\parallel}(t) \\ S_{sp,\parallel}(t) \end{bmatrix} \qquad (6)$$

with the following initial conditions: $S_{sp,\parallel}(t = 0) = \nu\alpha$ and $S_{sh,\parallel}(t = 0) = 1 - \nu$ where $\alpha = \exp(-D_{sp}b) = 1$. Here, $b$ is the b-value and $\nu$ is the volume fraction of dendritic spines on the dendritic segment. $D_{sh,\parallel}$ is the parallel diffusivity along the spiny dendritic branch. After solving the differential system in the Eq.(6), the total signal can be computed as

$$S = S_{sh,\parallel} + S_{sp,\parallel} = P_1' \exp(-D_1' q^2 t) + P_1' \exp(-D_2' q^2 t) \qquad (7)$$

where $P_1', P_2', D_1',$ and $D_2'$ are functions of $q$, given by

$$D_{1,2}' = \frac{1}{2}\left( X_{sh}' + X_{sp}' \mp \sqrt{\left(X_{sp}' - X_{sh}'\right)^2 + \frac{4}{q^4 \tau_{sh\rightarrow sp}\tau_{sp\rightarrow sh}}} \right)$$

$$X_{sh}' = D_{sh,\parallel} + \frac{1}{q^2 \tau_{sh\rightarrow sp}}; \ X_{sp}' = \frac{1}{q^2 \tau_{sp\rightarrow sh}}$$

$$P_1' = \frac{D_2'\left(1 - \nu(1-\alpha)\right) - (1-\nu)D_{sh,\parallel}}{D_2' - D_1'}$$

$$P_2' = \frac{(1-\nu)D_{sh} - D_1'\left(1 - \nu(1-\alpha)\right)}{D_2' - D_1'}$$

## 2.5 Extended Kärger Model to characterize both diffusion-mediated exchange between dendritic shaft and spines and permeative exchange

A recent study conducted in vitro experiments to investigate a three-compartment exchange system, incorporating both transmembrane exchange between intra and extracellular compartments and diffusion-mediated exchange within the intracellular space, mimicking the exchange mechanisms in GM[101]. However, a biophysical model incorporating both the contribution of diffusion-mediated exchange between the shaft and its dendritic spines and permeative exchange across the cell membrane is still lacking. Here, we propose an extended Kärger model as an attempt to fill this gap (see **Figure 2**). This model assumes: i) the diffusion in the dendritic spine is fully restricted (a dot like compartment), ii) the spine neck can be neglected; iii) the shaft can be modelled as sticks; and iv) extracellular diffusion can be modelled by isotropic gaussian diffusion. All compartments exchange with each other, with rates given by their corresponding exchange times: $\tau_{sp \to sh}$ and $\tau_{sh \to sp}$ between the dendritic spine and the shaft; $\tau_{sh \to e}$ and $\tau_{e \to sh}$ between dendritic shaft and the extracellular space; and $\tau_{sp \to e}$ and $\tau_{e \to sp}$ between the spine and the extracellular space. The extended Kärger system is then described by the following differential equation system:

$$\frac{dS_{sp}}{dt} = -D_{sp}q^2 S_{sp} - (\frac{1}{\tau_{sp \to sh}} + \frac{1}{\tau_{sp \to e}}) S_{sp} + \frac{S_e}{\tau_{e \to sp}} + \frac{S_{sh}}{\tau_{sh \to sp}} \qquad (8a)$$

$$\frac{dS_{sh}}{dt} = -D_{sh,\parallel}q^2 S_{sh} - (\frac{1}{\tau_{sh \to sp}} + \frac{1}{\tau_{sh \to e}}) S_{sh} + \frac{S_e}{\tau_{e \to sh}}S_e + \frac{S_{sp}}{\tau_{sh \to sp}} \qquad (8b)$$

$$\frac{dS_e}{dt} = -D_e q^2 S_e - (\frac{1}{\tau_{e \to sh}} + \frac{1}{\tau_{e \to sp}})S_e + \frac{S_{sp}}{\tau_{sp \to e}} + \frac{S_{sh}}{\tau_{sh \to e}} \qquad (8c)$$

Here $D_e$, $D_{sh,\parallel}$ and $D_{sp}$ are the corresponding diffusivities in the extracellular, axial diffusivity along the dendritic shaft (with $D_{sh,\parallel} = D_{sh}(\vec{g} \cdot \vec{n})^2$) and dendritic spines (assuming $D_{sp} = 0$). $S_e$, $S_{sh}$ and $S_{sp}$ are the diffusion signals for each compartment: extracellular, dendritic shaft and dendritic spine, respectively. The corresponding exchange rates have the following relations:

$$\frac{fv}{\tau_{sp \to e}} + \frac{f(1-v)}{\tau_{sh \to e}} = (1-f) * \left( \frac{1}{\tau_{e \to sh}} + \frac{1}{\tau_{e \to sp}} \right) \qquad (9a)$$

$$\tau_{sp \to sh} = \frac{v}{(1-v)}\tau_{sh \to sp} \qquad (9b)$$

$$\tau_{sp \to e} = \frac{vf}{(1-f)}\tau_{e \to sp} \qquad (9c)$$

$$\tau_{sh \to eh} = \frac{(1-v)f}{(1-f)}\tau_{e \to sh} \qquad (9d)$$

The parameters $f$ and $v$ are the volume fractions of intracellular and the dendritic spine compartments, respectively. We solved the differential system (Eqs.(8a)) numerically to obtain the directional total diffusion signal $\mathcal{K} = S_e + S_{sp} + S_{sh}$ and then computed numerically the corresponding direction-averaged signal as in Eq.(5).

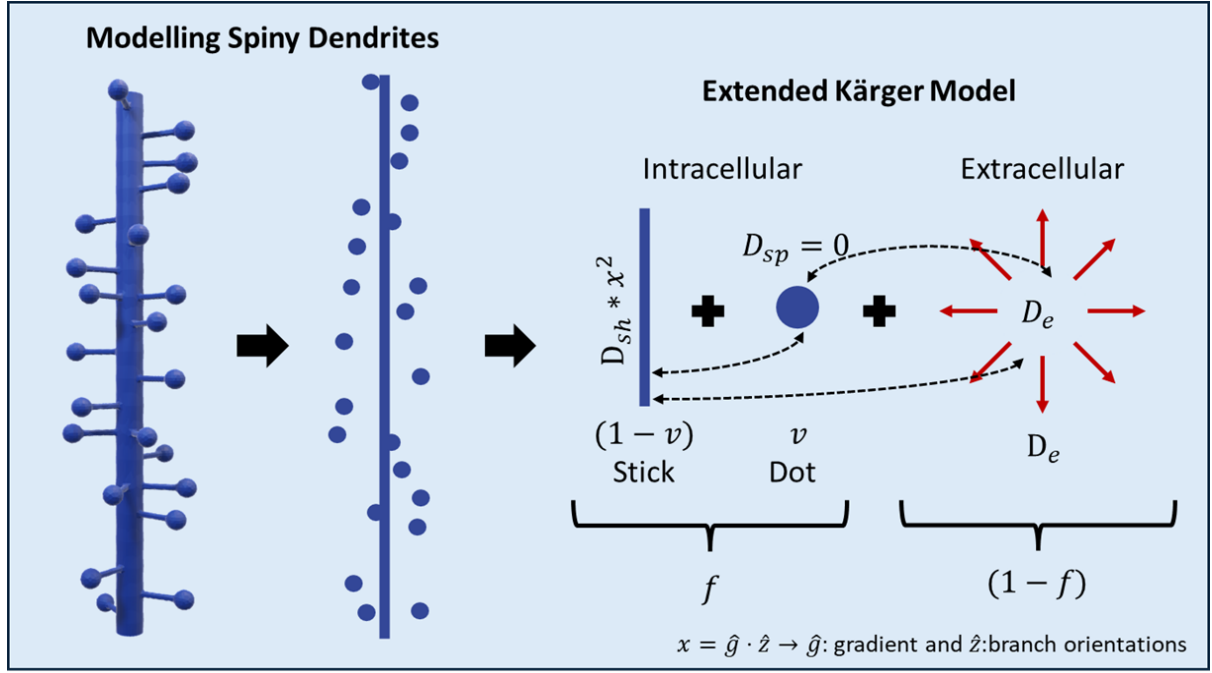

**Figure 2**: *Schematic of extended Kärger model. An additional compartment introduced to accommodate dendritic spines with diffusivity $D_{sp} = 0$. The dendritic shaft is modelled as a stick compartment and extracellular space is assumed to be isotropic gaussian compartment with diffusivity $D_e = 1.2 \frac{\mu m^2}{ms}$. All three compartments exchange with each other at specific rates.*

# 3  Methods

In this section, our objective is to design virtual experiments of time-dependent SDE measurements using Monte Carlo simulations to investigate the impact of diffusion-mediated exchange between dendritic shaft and spines and compare our results with theoretical predictions from the previous section.

We first describe how we built digital substrates representative of spiny dendrites and then we describe our simulation experiments and analysis.

## 3.1  Spiny Dendrite Branches

To understand the diffusion-mediated exchange between the dendritic shaft and its spines, we ran simulations in digital models built either from 3D electron microscopy or by geometrical modelling. The former is more realistic, while the latter is more flexible. The two digital models are described below.

### 3.1.1  3D-Reconstructed spiny dendrites

A dataset consisting of 3D reconstruction of pyramidal cells from layer 2-3 of a P36 mouse's visual cortex, obtained from electron-microscopy data was downloaded from MICrONs explorer[102–104].

Using "trimesh" package in Python, the surface mesh dataset was initially converted into PLY format to extract spiny dendritic branches in Blender v4.1[105]. 13 spiny dendritic branches were extracted from the 3D reconstructed pyramidal cells. The length of the dendritic branches varied from 30 to 110 μm. Using object modifiers in Blender, the mesh quality was improved and cleaned from residual artifacts for each dendritic branch. An exemplary branch is depicted in **Figure 3A** before and after mesh processing. To quantify the spine morphometry and isolate the spins in the dendritic spines, each dendritic branch was manually cleaned from spines as shown in **Figure 3A**. The morphological details of real spiny branches are documented in the chart (see **Figure 3B**); the total volume and surface area of real dendritic branches were compared with their corresponding dendritic shaft morphometry. Accordingly, the mean spine volume fractions and surface area fractions in real branches were estimated to be $0.41\pm0.11$ μm$^3$ and $0.55\pm0.06$ μm$^2$ (mean $\pm$ standard deviation), respectively. Furthermore, the mean spine density was estimated to be $1.17\pm0.06$ μm$^{-1}$ (mean $\pm$ standard deviation). All dendritic branches were aligned in the $\hat{z}$ direction to simplify the Monte Carlo simulations.

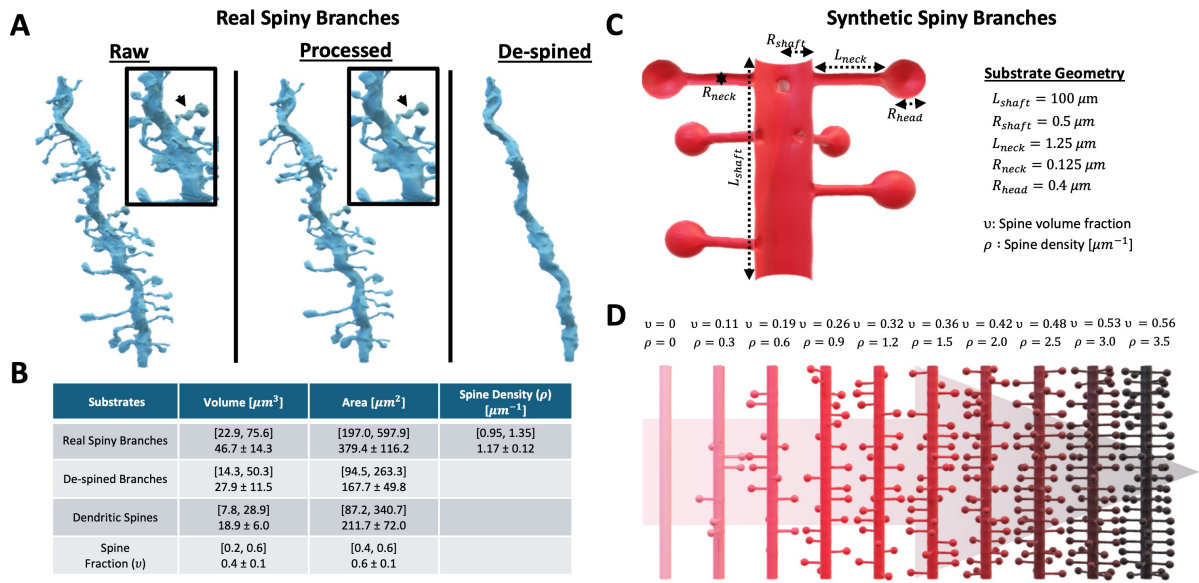

**Figure 3**: *(A) illustration of real dendritic branches with spines from electron microscopy 3D segmentation. The substrates were processed for quality improvement and cleaning the sharp edges and mesh artifacts. After the mesh processing, the spiny branches were de-spined for isolating the particles inside the spines. (B) Morphological information of the real dendritic branch segmentations; minimum and maximum values of volume, surface area and spine densities are given in the brackets together with corresponding mean and standard deviation values. (C) An exemplary toy model for spiny dendritic branches. The substrate geometry was tuned to match real dendritic spine morphology (volume and length) and mimic the particle dynamics in the real branches. (D) Toy substrate models for dendritic branches are shown for increasing spine densities (ρ) and the corresponding spine volume fractions (ν) are also documented.*

### 3.1.2   Digital model for spiny dendrites

We propose a simple tunable digital model of spiny branches for understanding the diffusion-mediated exchange mechanism within intracellular compartments (i.e. between dendritic spines and dendritic shaft). The digital model represents dendritic spines with mushroom-like morphology, since they are the most mature and stable type[106,107]. We model dendrites and their spines by spherical spine heads connected to a cylindrical dendritic shaft through a narrow cylindrical neck. The dendritic shaft is aligned in the $\hat{z}$ direction and all spines are attached to the dendritic shaft perpendicularly. The spines are placed at nodes randomly arranged along the dendritic shaft. To avoid overlapping of multiple spines at the same node, we imposed an angular separation of 90° in the plane perpendicular to the shaft between the neck of the current spine and the neck of the next one, after the first randomized spine placement. For spine densities exceeding ~1.2 μm$^{-1}$, complete randomness was unattainable with our approach, leading to regular ordered arrangement. To assess the impact of spine arrangement, in preliminary experiments, we analyzed the diffusivity and kurtosis time-dependence along the main dendritic axis. At spine densities ≤1.2 μm$^{-1}$, we observed signatures of both 1D short-range disorder and diffusion-mediated exchange, as expected[38,108,109]. However, at higher densities, only diffusion-mediated exchange was observed. It is worth noting that all the spiny branches from real 3D reconstructions used here had spine density ≤1.2 μm$^{-1}$, and that regular ordered spine arrangements along dendrites is not unrealistic – and in fact it is common - for neurons with high spine density. For example, in Purkinje cells (where spine density ≥2 μm$^{-1}$), spines are arranged in helical patterns, positioned regularly along the dendrite with constant spacing (~0.5-0.6 μm) and angular displacement between them[110].

Skeletons of spiny dendritic branches were built on MathWorks MATLAB 2022a[111] involving functions from the Trees-Toolbox[112]. The generated skeletons were exported as SWC file format for surface meshing in Blender API v4.1 using metaballs and the Blender addon publicly available at: https://github.com/kdrsimsek/SWC_Mesher_v4x. The metaballs employed in 3D reconstruction of substrates were scaled to match the desired morphological sizes in each substrate. The 3D reconstructed spiny dendritic branches were further processed for cleaning from any mesh artifacts. Due to surface meshing technique, the nominal sizes in the dendritic branches are slightly different from the prescribed skeleton ones. The dendritic spine sizes were based on the average values reported in the literature (see section 2.1). The sizes of the dendritic shaft were defined as 100 μm and 0.5 μm for shaft length $L_{shaft}$ and shaft radius $R_{shaft}$, respectively. The resulting substrate models for the spiny dendritic branches are illustrated in detail in **Figure 3C-D**, providing morphological information on the substrates.

The spine morphology in the simple digital models were tuned to have similar particle dynamics as in the case of real spiny dendritic branches, but without the variability in morphology between spines, to simplify the problem. We systematically optimized our synthetic spiny dendritic substrates by adjusting the spines' morphological parameters (spine head radius $R_{head}$, and neck length $L_{neck}$) to reproduce the particle escape dynamics observed in real dendritic spiny branches. We maintained the volume of each spine in our synthetic model equal to the average volume of the spines from real dendrites; however, we note that this approach does not necessarily equate to matching spine morphology. Then, the best matching configuration for the spine geometry was determined for the final digital substrate design. **Figure 4** depicts the matching correspondence of tunable substrates to the real spiny dendritic substrates. Accordingly, the spine head radius $R_{head}$ was set to 0.4 µm and spine neck sizes were defined as 1.5 µm and 0.125 µm for neck length $L_{neck}$ and radius $R_{neck}$, respectively.

After optimizing our digital models, two sets of spiny dendritic segments were generated to investigate the exchange effect of diffusion in spines in the case of simple straight shaft and in the case of shaft with additional complexity, introducing undulation and/or beading. Our aim in generating the first set of branches was to study the impact of different spine densities (and total spine volume fraction) on the diffusion exchange mechanism, without additional confounders stemming from shaft undulation and beading. Therefore, ten spiny branches with typical spine densities ($\rho = [0 - 3.5]$ µm$^{-1}$) were generated[113,114]. The second set of spiny dendritic segments was generated to study the contributions stemming from other morphological features, i.e., beading and/or undulation. A total of 32 spiny branches at $\rho = [1, 2]$ µm$^{-1}$ were generated using the combination of four beading amplitudes ($A_{bead} = [0,1,2,3]$ µm) and four undulation periods ($N_{und} = [0, 0.75, 1.5, 3]$, yielding wavelengths $\lambda_{und} = [0,30,60,120]$ µm). Undulations were generated by cosine function along $\hat{x}$ axis and beadings were generated by a smoothed uniform sampling between $2R_{shaft}$ and $2R_{shaft} + A_{bead}$. In the beaded branches, the spine neck lengths $L_{neck}$ were maintained to be around 1.25 µm by accounting for the beading amplitudes. Notably, any of these features can be changed arbitrarily, here, we investigated realistic spine densities and realistic to extreme undulations and beading in dendritic branches. All generated digital substrates and the codes for data analysis will be available on https://github.com/kdrsimsek upon publication.

## 3.2   Monte Carlo Diffusion Simulations

The Python library disimpy[115] was used for Monte Carlo diffusion simulations. Two types of simulations were performed in this study, with the diffusivity set to the typical value for intra-neurite water molecules[5,13], 2 $\frac{µm^2}{ms}$, and periodic boundary conditions used for all diffusion simulations. The number of spins $N_s$ and the number of time steps $N_t$ in all simulations were determined by the Monte Carlo convergence method[116,117]. The first set was used to generate particle trajectories, which were then utilized to understand particle dynamics in both real and synthetic spiny branches, and to tune the synthetic branches based on the dynamics observed in the real spiny branches. In this set, $N_s = 10^4$ spins were simulated with a time step of 20 μs to record particle trajectories. The spins were initialized separately in the dendritic spines and the shaft to analyze their individual contributions to the particle dynamics. The time dependence of the number of spins inside dendritic spines (the concentration curve $\eta$) was computed using the diffusion trajectories obtained from Monte Carlo simulations. For each real dendritic branch, $\eta(t)$ was determined, excluding any particles exiting the dendritic spines. To eliminate bias due to particle density within the dendritic spines, the initial number of particles at $t = 0$ was set to be $N_s = 10^4$ for each substrate. Consequently, the normalized $\eta(t)$s characterized the particle escape dynamics and were compared to those computed from the real spiny dendritic branches (**Figure 4**).

After optimizing the synthetic substrates to mimic the particle dynamics of real dendrite segments, the second set of simulations was configured to have $N_s = 10^6$ spins and $N_t = 10^4$ time steps to generate diffusion signals. These simulations were designed to investigate the impact of dendritic spines on the diffusion exchange mechanism. Ten linearly spaced b-values up to $7 \frac{ms}{\mu m^2}$ were employed, with 128 evenly distributed directions for direction averaged signal, including parallel ($\hat{z}$) and perpendicular ($\hat{x}$ and $\hat{y}$) to the dendritic shaft (oriented along $\hat{z}$) for the signals in parallel ($S_\parallel$) and perpendicular ($S_\perp$) directions[46,47]. The maximum b value was determined empirically, selected to ensure sufficient diffusion-weighted signal attenuation across all examined spine densities. A total of 15 different schemes were constructed by combining eight diffusion gradient separation times ($\Delta = [10,25,35,55,85,160,235,310]$ ms), with two gradient durations ($\delta = [3,15]$ ms, i.e. narrow and wide gradient pulses, respectively) which were tailored for preclinical scanners employing ultra-strong gradients (i.e. $\geq 300$ mT/m).

## 3.3 Data Analysis

### 3.3.1 Diffusion and exchange analysis in dendritic spines
First, we examined whether the modified two-compartment Kärger model (Eq.(6)) could accurately describe $S_\parallel$. We fitted Eq.(7) to estimate the residence times $\tau_{sp \to sh}$ and $\tau_{sh \to sp}$ and compare them with the theoretical predictions from Eq.(1) and Eq.(3)

Then, we investigated numerically the impact of spine morphology on the exchange time estimates using theoretical predictions from the analytical solutions (Eq.(1) and Eq.(3)). For the $\tau_{sp \to sh}$ analysis, uniform distributions of spine head radius $R_{head}$ (mean: 0.5 μm, standard deviation: 0.2 μm), and neck length $L_{neck}$ (mean: 1.25 μm, standard deviation: 0.5 μm) were generated while the neck radius $R_{neck}$ was kept at 0.125 μm. Similarly, for the $\tau_{sh \to sp}$ analysis, uniform distributions of neck radii $R_{neck}$ (mean: 0.175 μm, standard deviation: 0.07 μm) and shaft radii $R_{shaft}$ (mean: 0.55 μm, standard deviation: 0.26 μm) were generated.

In addition to dendritic spine morphology, we also inspected the influence of dendritic shaft morphology by using synthetic spiny substrates with varying levels of undulation and beading. We estimated the exchange times for these different undulation and beading levels to explore the impact of dendritic shaft morphology on exchange time estimations.

### 3.3.2 Impact of dendritic spines on time-dependent SDE measurements and NEXI/SMEX analysis

To evaluate how diffusion-mediated exchange between dendritic shaft and spines influences exchange time estimates in NEXI/SMEX analysis, we conducted two experiments. First, we simulated time-dependent SDE signals using our synthetic substrates of spiny dendrites in extracellular space (with impermeable membranes) and fitted NEXI/SMEX to assess how spine density affects permeative exchange time estimates. Second, we incorporated membrane permeability using our extended three-compartment Kärger model to examine how the combined effects of permeative and diffusion-mediated exchange alter these estimates.

For the first experiment, to replicate diffusion signal from a dMRI voxel and compare it with NEXI/SMEX biophysical modelling of in-vivo experiments, we simulated the total signal $S_{total}$ arising from a dMRI voxel as the weighted sum of intra and extraneurite signals, $S_{neurite}$ and $S_{Gaussian}$, respectively: $S_{total} = 0.75 \times S_{neurite} + 0.25 \times S_{Gaussian}$. The $S_{neurite}$ was derived from our simulations in spiny dendrites at different spine densities (with and without undulations and/or beadings) while $S_{Gaussian}$ is mono-exponential decay with diffusivity of $1.2 \frac{\mu m^2}{ms}$. $S_{total}$ was fitted using the NEXI model to estimate the exchange time ($t_{ex}$) for all simulated conditions.

For the second experiment, to investigate the contributions of exchange mechanisms (i.e. dendritic spines and permeative exchange), we used numerical solutions of the three compartment Kärger model defined in Section 2.5 to generate diffusion signals, incorporating characteristic membrane permeability values ($\xi = [1 - 20] \times 10^{-3} \frac{\mu m}{ms}$)[14,118] and typical total spine volume fractions ($\nu \leq 20\%$). Then, to predict the signal from a millimeter-sized voxel, we calculate the powder-averaged signal computing numerically the integral over the angle ($\theta$) between the diffusion-sensitizing gradient direction and neurites bundle direction: $\int_0^1 S(q, t, ; \theta) d(\cos \theta)$. The characteristic exchange time from spine-to-shaft ($\tau_{sp \to sh}$) was set to be the average value predicted according to the heterogeneity of spines morphology (**Figure 7**);

the characteristic exchange times from shaft to extracellular space ($\tau_{sh \to e}$) and from spine to extracellular space ($\tau_{sp \to e}$) were computed by using the relation, $\tau = \frac{1}{SVR\xi}$, where SVR is the surface-to-volume ratio of the shaft and spine, respectively. The diffusivities of each compartment were $D_{sp} = 0 \frac{\mu m^2}{ms}$ for spines (i.e., still water or dot compartment), $D_{in} = 2.5 \frac{\mu m^2}{ms}$ for the intra-neurites and $D_e = 1.2 \frac{\mu m^2}{ms}$ isotropic Gaussian extracellular space. The corresponding compartment signal fractions were assumed to be 75% intracellular and 25% extracellular space. Then, we solved numerically the system of differential equations Eqs.(8a) for the extended Kärger model to quantify the impact of membrane permeability and dendritic spines. Later, we utilized the NEXI model to estimate the exchange time of the system to characterize the impact of diffusion-mediated exchange and permeative exchange across cellular membrane.

# 4 Results

## 4.1 Particle exchange dynamics

The proposed tunable system for synthetic spiny dendrites accurately reflects the particle dynamics in real spiny dendrites. The characteristics of the normalized concentration curves $\eta(t)$ extracted from real and synthetic substrates are compared in **Figure 4A**, exhibiting similar particle dynamics. In **Figure 4B**, all normalized $\eta(t)$ values of synthetic spiny branches at varying spine densities are compared with those of real spiny dendritic branches. The tunable synthetic substrates demonstrated accurate particle dynamics, falling within the range of variation caused by morphological differences in real spiny branches.

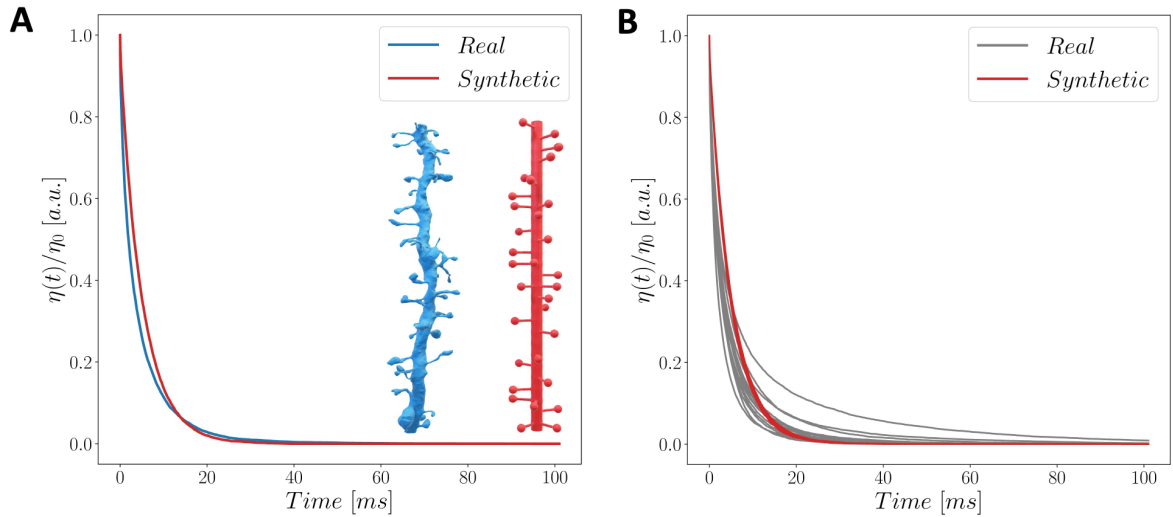

**Figure 4**: *The comparison of particle diffusion dynamics in the dendritic spines; between (A) one to one comparison and (B) all real and synthetic branches. η describes the number of particles in the dendritic spines as a function of time, normalized by the initial number of particles ($\eta_0$ =10000). In addition to morphological complexity and variety in the real dendritic spines, the particles' first exit rates from spine-to-shaft present similar characteristic decays in the overall comparison between each group of substrates (real vs synthetic).*

## 4.2    Diffusion & exchange mechanism in spiny dendrites

The results of the fitting of the modified two-compartments Kärger model to the MC simulated diffusion-weighted signals parallel to the shaft main axis ($S_\parallel$) are reported in **Figure 5**. In **Figure 5A,** the predicted signals from the fitted modified Kärger model are overlayed on the simulated signals ($S_\parallel$) at all diffusion times for the narrow pulse scheme. **Figure 5B** illustrates the estimated (dashed lines) spine-to-shaft residence time $\tau_{sp \to sh}$ and the total exchange time of the system $t_{ex}^{DM}$ as a function of spines density and compares them with the corresponding theoretical predictions (solid lines). Both estimated times match well the theoretical predictions. Similar results for wide gradient pulse scheme are reported in **Supplementary Figure S1**. However, for the wide pulse condition, the estimated exchange times are shorter than the theoretical results.

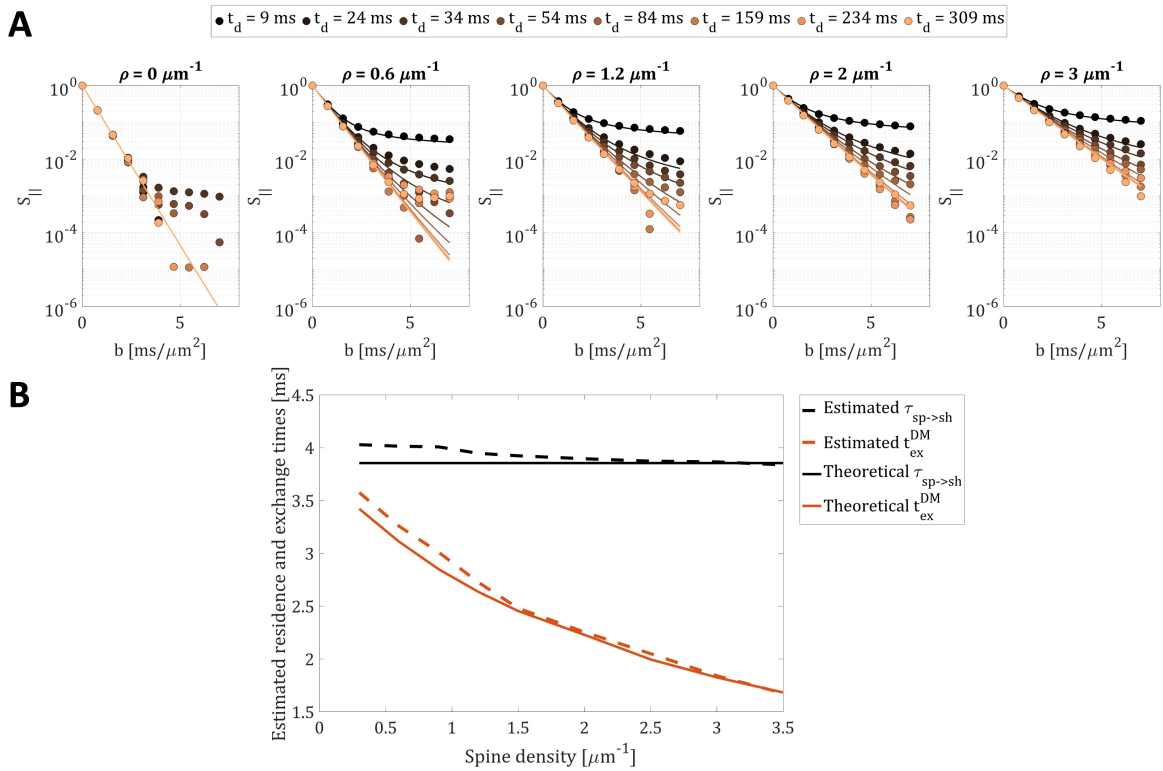

**Figure 5**: *Theoretical signal predictions from the modified Kärger model are overlaid on simulated parallel diffusion signals ($S_\parallel$) across all diffusion times using the narrow pulse scheme (A). Additionally, model predictions are shown over a range of spine densities ($\rho$), illustrating the impact of $\rho$ on diffusion behavior. Here we modelled the signal decay parallel to a given spiny dendrite using the modified Kärger model proposed in section 2.4. Briefly, two exchanging compartments (shaft and spines), where one compartment is modelled as fully restricted diffusion (the spines modelled as gaussian compartment with diffusivity zero). (B) The panel shows the estimated residence time $\tau_{sp \to sh}$ (black) and exchange time $t_{ex}$ (orange) obtained using the modified Kärger model, under the narrow gradient pulse settings. These estimates are compared with the corresponding theoretical predictions ($\tau_{sp \to sh}$ and $t_{ex}^{DM}$, see section 2.2). The dashed lines present the estimations of the modified Kärger model while the theoretical predictions are displayed as solid lines for both spine-to-shaft residence times and exchange times. Both residence and exchange time estimates align with the theoretical predictions very well.*

As a supplementary analysis, we also investigated the time-dependence of the apparent diffusion ($ADC$) and kurtosis ($K$) coefficients parallel to the main dendritic axis (details in Supplementary Materials) and the results, reported in **Supplementary Figure S2**, show that both $ADC$ and $K$ decrease with time up to a diffusion time $t_d \leq 54$ ms, then only $K$ continue to decrease monotonically up to $t_d \leq 154$ ms, indicating signature of barrier-limited exchange.

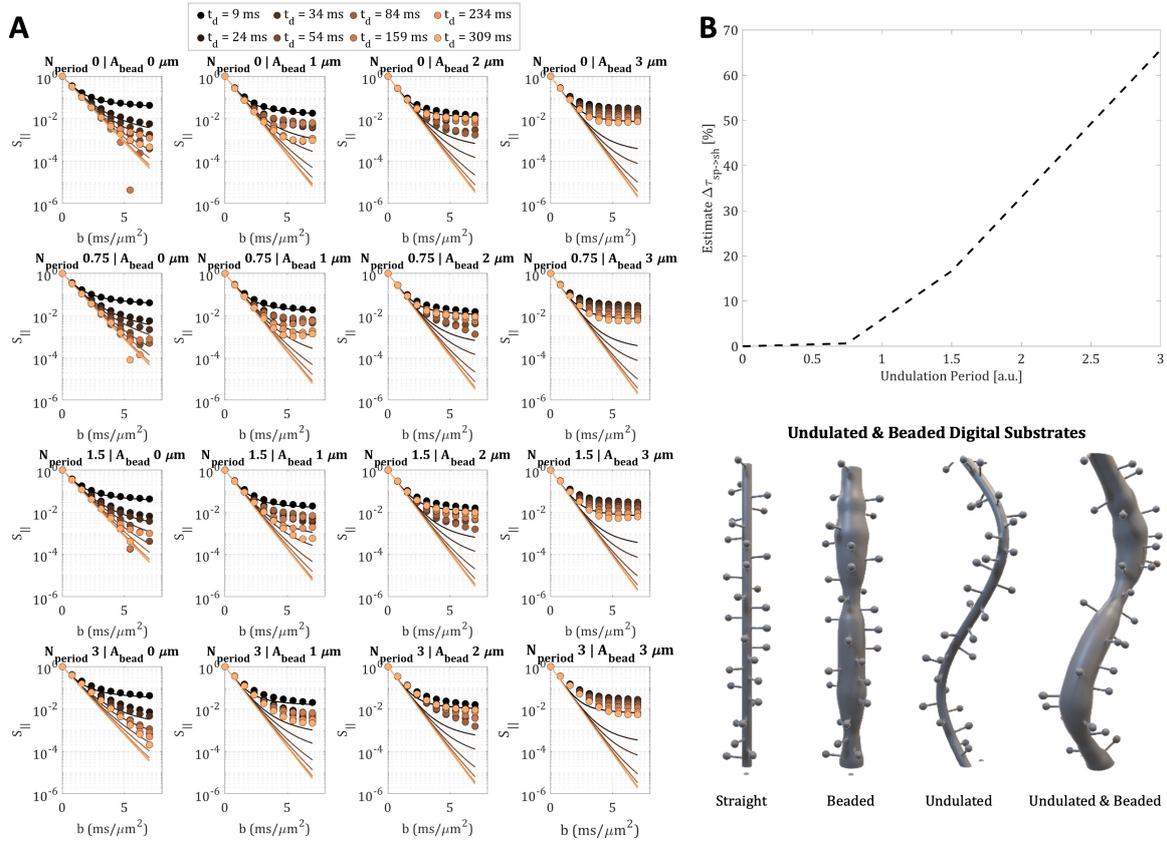

***Figure 6***: *The diffusion signals in the parallel direction obtained from undulated and beaded synthetic substrates with spine density $\rho = 1\ \mu m^{-1}$ are shown in the figure. The diffusion signals generated using narrow pulse gradient scheme are presented in (A). At long diffusion times ($t_d > 150\ ms$), the restriction effect starts dominating the diffusion signal. In (B), the percentage change in the spine-to-shaft residence times ($\Delta \tau_{sp \to sh}$), estimated by fitting diffusion signals to the modified Kärger model, is shown as a function of undulation period $N_{und}$, relative to a reference branch without undulation or beading. For undulation per $N_{und} < 1.5$, the $\tau_{sp \to sh}$ results present no impact of undulation on the estimated residence times.*

The simulated diffusion signals in parallel direction within undulated and beaded synthetic substrates at spine density of $\rho = 1$ μm$^{-1}$ are shown in **Figure 6A** for the narrow gradient pulse scheme (see **Supplementary Figure S4** for wide gradient pulses). We fitted the modified Kärger model to the simulated signals to estimate $\tau_{sp \to sh}$ and computed the percentage change $\Delta\tau_{sp \to sh}$ relative to the estimates obtained from a substrate without undulation and beading ($\tau_{sp \to sh} \simeq$ 4 ms). $\Delta\tau_{sp \to sh}$ as function of undulation period $N_{und}$ is reported in **Figure 6B**. No impact of undulation on $\tau_{sp \to sh}$ is observed for realistic undulation levels ($\Delta\tau_{sp \to sh} \sim 0\%$ for $N_{und} < 1.5$)., However, for undulation periods $N_{und} \geq 1.5$, a significant overestimation (up to 65%) is observed in the estimated $\tau_{sp \to sh}$ results. We only reported the impact of undulation as the modified Kärger model fits well only these signals, while it cannot characterize well the signals from substrates with beading. Since beading introduces additional restriction to the diffusion, making the signal restriction-dominated and leading to a mismatch with the model. We also note that at long diffusion times ($t_d \geq 150$ ms), the effects of restriction become dominant in the diffusion signal, which deviates from the model predictions at high b values.

## 4.3 Impact of dendritic spine morphology and shaft size on exchange times

The histograms of $\tau_{sp \to sh}$ and $\tau_{sh \to sp}$ obtained using Eq.(1) and (3) for the typical values of spine and shaft morphologies at spine density 1.2 μm$^{-1}$ in the healthy adult brain are shown in **Figure 7A-B**, respectively. The morphological analysis determined the expected average $\tau_{sp \to sh}$ to be approximately 10 ms and the volume-weighted average $\tau_{sp \to sh}$ to be approximately 26 ms **Figure 7C** illustrates the impact of spine morphology on $\tau_{sp \to sh}$, based on the prediction of Eq.(1). The isocurves highlight degeneracy: the same residence times can result from very different spine geometries (i.e. neck length $L_{neck}$ and spine head radius $R_{head}$). For instance, particles in the filopodia-like or mushroom-like spines can have the same residence time. Determining the spine morphology from the exchange estimations alone is thus a degenerate problem. Likewise, different shaft ($R_{shaft}$) and spine neck ($R_{neck}$) radii can lead to same $\tau_{sh \to sp}$ values (**Figure 7D**).

Worth noting that the $\tau_{sh \to sp}$ values estimated from our Monte Carlo simulations are a few orders of magnitude higher than the theoretical predictions from Eq.(3) as shown in **Supplementary Figure S3A**. We further investigated this discrepancy by testing it for multiple neck sizes ($R_{neck} =$ [0.125, 0.250, 0.375] μm) and quantified the difference between simulated and theoretical results using linear regression (**Supplementary Figure S3B**). To correct for such discrepancy, we then applied the estimated scaling factor to the $\tau_{sh \to sp}$ predicted by Eq.(3) to obtain the "corrected" values reported in **Figure 7B-D**.

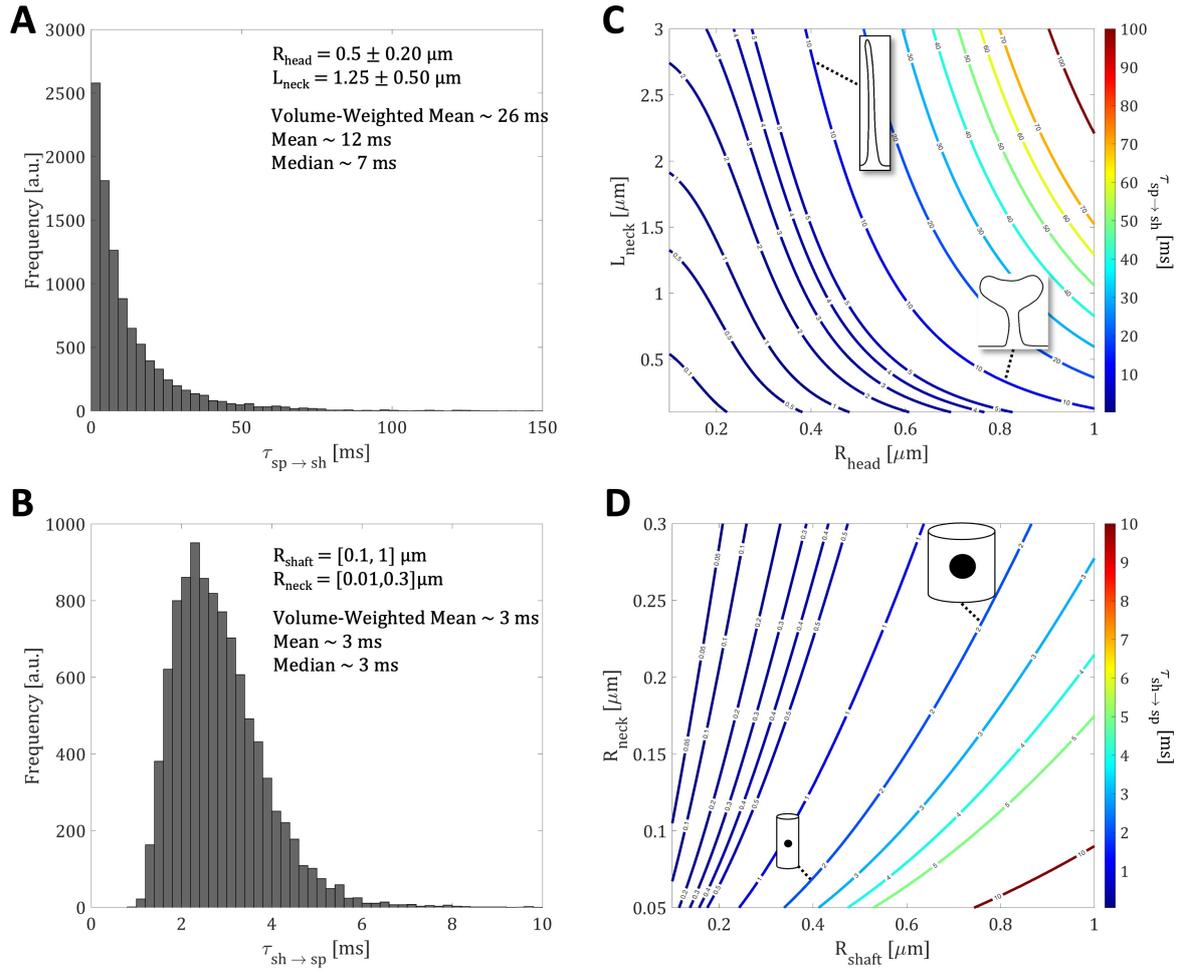

***Figure 7***: *(A, B) Theoretical prediction of the distribution of spine→shaft and shaft→spine residence times for a distribution of heterogeneous spine and shaft morphologies. (C, D) Theoretical prediction of spine→shaft and shaft→spine exchange times versus spine neck length and head radius (for spine→shaft) and shaft radius and spine neck length or neck radius (for shaft→spine) to show the degeneracy and inability to differentiate spine and shaft morphology from residence time.*

## 4.4 Impact of dendritic spines on time-dependent SDE measurements and NEXI/SMEX analysis

**Figure 8** illustrates the diffusion-mediated exchange system for the synthetic substrates, incorporating a Gaussian extracellular compartment with a volume fraction of 0.25. The estimated exchange times ($t_{ex}^{NEXI}$) obtained from fitting the NEXI model to the total diffusion signal ($S_{total} = 0.75 \times S_{neurite} + 0.25 \times S_{Gaussian}$) are illustrated as a function of total spine volume fraction for narrow gradient pulse scheme (see **Supplementary Figure S6** for the results with the wide gradient pulse). In the narrow pulse simulation results, the estimated exchange times $t_{ex}^{NEXI}$ are between 10-150 ms range for typical total dendritic spine volume fractions ($\nu \leq 20\%$)[118]. These values are of the same order of magnitude of those estimated in-vivo in the cerebral cortex of healthy adult human and mouse brains using similar SDE protocols with diffusion time $t_d \leq 65$ ms and gradient duration $\delta = [4.5 - 17.5]$ ms [8,14,15,35,36]. However, the $t_{ex}^{NEXI}$ estimations in the wide pulse acquisitions are drastically higher ($t_{ex}^{NEXI} > 200$ ms).

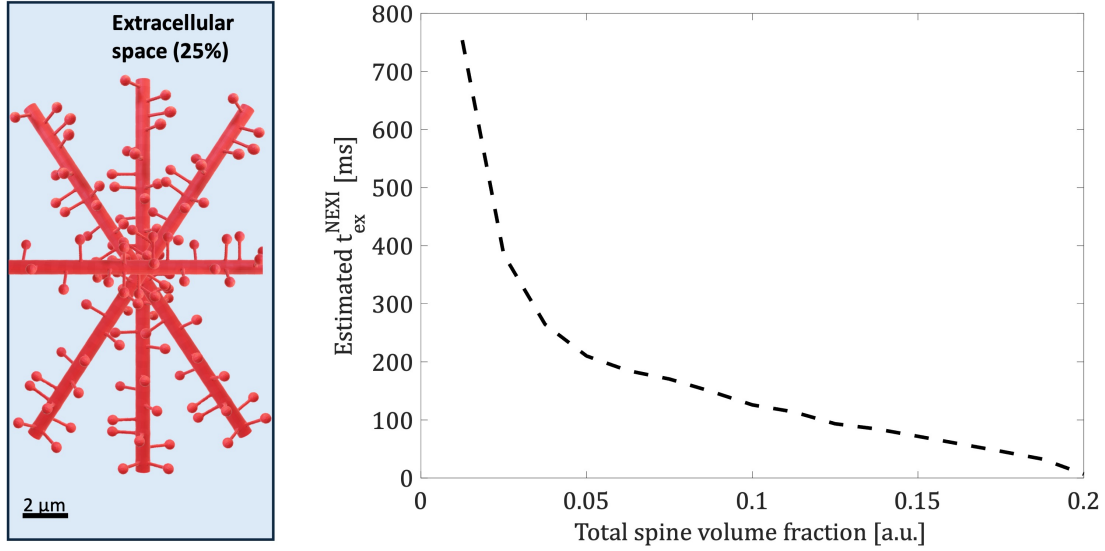

**Figure 8**: *Bias in NEXI estimates of $t_{ex}^{NEXI}$ when spines and exchange with spines are not considered, for narrow gradient schemes. We generated the signal using 75% spiny dendrite + 25% isotropic gaussian extracellular space with diffusivity $1.2 \frac{\mu m^2}{ms}$ and fitted the NEXI model to it for different spine densities. The results (dashed line) are smoothed with a one-point window for better visualization.*

To assess the impact of both exchange mechanisms (i.e. permeative and diffusion-mediated exchanges) on the exchange time measurements, **Figure 9** presents a sensitivity analysis of exchange times as a function of membrane permeability and total spine volume fraction. As an illustration, **Figure 9A** depicts the fitting of the NEXI/SMEX model to diffusion signals simulated at multiple diffusion times ($t_d \leq 54$ ms) with a membrane permeability of $\xi = 10^{-2} \frac{\mu m}{ms}$ and a total spine volume fraction of $\nu = 0.1$. The estimated exchange times ($t_{ex}^{NEXI}$) as a function of permeability and total spine volume fraction are shown in log scale in **Figure 9B**. The exchange times are smoothed with a gaussian filter and presented in the $\log_{10}$ scale for better visualization and differentiation of degeneracy bands. The estimated exchange times in our simulations vary between $1 - 100$ ms and decrease with both increasing permeability and total spine volume fraction. The degeneracy bands are highlighted with isocurves (black lines) and reporting the corresponding exchange time in milliseconds. Furthermore, the percentage change of $t_{ex}^{NEXI}$ estimations are presented as a function of permeability (with $\nu = [0.04, 0.2]$) and total spine volume fraction (with $\xi = [6 - 18] \times 10^{-3} \frac{\mu m}{ms}$) in **Figure 9C**, and **D**, respectively. In panel C, we show that membrane permeability contributes to up to a 90% reduction in the $t_{ex}^{NEXI}$ estimates, while in panel D, the total spine volume fraction accounts for up to an 80% reduction. However, these effects cannot be directly attributed to each mechanism individually, as their influences are interdependent. For instance, the higher $\nu$ values result in faster diffusion-mediated exchange as well as permeative exchange due to the higher SVR of dendritic branch. Therefore, our findings indicate that both increase in membrane permeability and/or spine density can lead to reduction of the estimated $t_{ex}^{NEXI}$. The disentangling of the contributions of membrane permeability and presence of dendritic spines to the exchange mechanisms is not trivial as well as degenerate when using time-dependent SDE measurements.

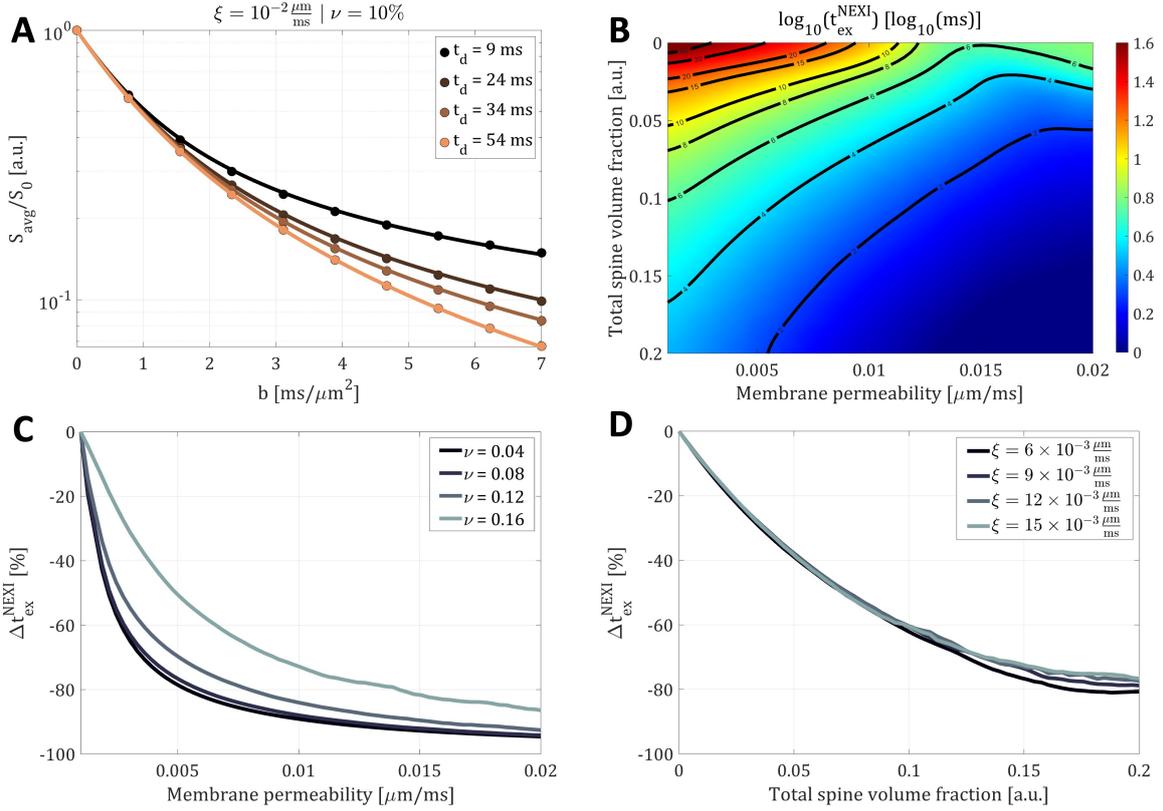

**Figure 9**: (A) Simulated diffusion signals at different diffusion times ($t_d$) using three-compartment extended Kärger model (data points) (spine, shaft and extracellular space) are fitted with the NEXI/SMEX model (solid lines) for given membrane permeability ($\xi = 10^{-2}\frac{\mu m}{ms}$) and total spine volume fraction ($\nu = 0.1$). The NEXI/SMEX model perfectly fits the generated extended Kärger model signals. (B) The estimated exchange times ($t_{ex}^{NEXI}$) are mapped using logarithmic scale times as a function of the membrane permeability and total spine volume fraction. The black isocurves highlight the degeneracy in the exchange estimations and documents the $t_{ex}^{NEXI}$ value in ms. The exchange estimates decrease with both the total spine volume fraction and the membrane permeability. Panels (C) and (D) illustrate the percentage change in $t_{ex}^{NEXI}$ as a function of membrane permeability ($\xi$) for multiple total spine volume fractions ($\nu = [0.04, 0.16]$), and as a function of total spine volume fractions ($\nu$) for several membrane permeability values ($\xi = [6, 15] \times 10^{-3}\frac{\mu m}{ms}$), respectively. The panels quantify the contributions of $\xi$ and $\nu$ to the exchange estimations.

# 5 Discussion

This study explores the influence of dendritic spines on the diffusion exchange mechanism using MC simulations and quantifies their impact on water exchange time estimations from time-dependent SDE measurements. We propose to use the analytical solutions of the narrow escape problem to characterize diffusion-mediated exchange in spiny dendrites. Furthermore, our findings demonstrate that, even in non-permeative diffusion simulations, the presence of dendritic spines produces a time-dependent signature similar to that observed in permeative exchange. Here, we propose the dendritic spines as an alternative non-permeative exchange mechanism[16] that needs to be taken into account when interpreting exchange time estimates from time-dependent SDE measurements.

Anomalous diffusion in dendritic spines has been extensively studied in literature. Before diffusion MR, interest in molecular diffusion within dendritic spine was pioneered in cellular biology, leading to the development of various diffusion models to better understand these processes[85,86,93,119]. In the diffusion MR field, few studies focused on the impact of dendritic spines on the diffusion measurements. A study examined the metabolite diffusion in silico and in vivo in rodents to inspect the effects of dendritic spines on the diffusion signal using diffusion-weighted MR spectroscopy[120]. This work focused on restriction and showed that the presence of dendritic spines leads to larger estimates of neurite size and SVR from SDE data acquired with both pulsed gradients at high b value and oscillating gradients at high frequency. Following works further investigated the impact of dendritic spines on both metabolite and water diffusion signals[46–48,121] and provided evidence supporting the hypothesis that it is possible to measure the impact of dendritic spines on the intracellular diffusion MR signal. Only some recent studies focused on the potential role of dendritic spines in diffusion exchange measurements[46,48].

In this study, we present an in-depth analysis of the potential role of dendritic spines in diffusion exchange measurements using SDE, extending and complementing prior research. We evaluate how well different biophysical models - based on the popular Kärger model - describe diffusion-mediated exchange between dendritic shafts and spines, leveraging ad-hoc Monte Carlo simulations and theoretical insights from the 'narrow escape problem'. Furthermore, we assess the extent to which such diffusion-mediated exchange biases permeative exchange estimates in time-dependent SDE measurements when analyzed with NEXI/SMEX modeling. Our findings highlight the importance of accounting for diffusion-mediated exchange in SDE measurements, cautioning against interpretations based exclusively on permeative exchange mechanisms. Below, we discuss each of these contributions in detail.

## 5.1 Diffusion-mediated exchange in spiny dendrites

Adapting 'the narrow escape problem' to biophysical modelling of exchange in spiny dendrites seems feasible and agrees with the modified Kärger model of two exchanging compartments[12] (**Figure 5**). Considering the spine-to-shaft exchange, the diffusion model in dendritic spines[93], that characterizes the expected residence time of molecules within the spines, agrees very well with the exchange time estimations from the modified Kärger model[12] with narrow pulse settings (<8% difference between estimated and theoretically predicted times). In addition, one of our major findings shows that the modified Kärger model Eq.(7) accurately characterizes the diffusion signal within spiny dendrites and the corresponding intracellular time-dependent SDE signal (**Figure 5A**). However, it is worth mentioning that there is some bias in the fit with narrow pulses, at short diffusion times and low spine densities. This is probably due to noise effects, which are particularly important in those cases where the signal decays faster. The results obtained from the wide pulse shows underestimated exchange time with the modified Kärger model (**Supplementary Figure S1**). The Kärger model assumes that the impact of exchange during diffusion encoding is negligible, and the probability of spins' membrane permeation is constant, i.e. short gradient pulse approximation[1,12,97,98]. Accordingly, the wide pulses in the SDE measurements, violating the model assumption, lead to overestimation in residence time estimates in the model fitting[14,35,98]. Here, we raise caution when interpretating the model predictions obtained from wide pulse measurements.

The diffusion-mediated exchange between spine and shaft exhibits the time dependent signature of permeative exchange across cell membrane and yields an average exchange time on the order of 10 ms. This estimate aligns with previously reported permeative exchange times (10–60 ms) for neurites with radii in the range of 0.25–1 μm and membrane permeabilities $\xi = [1 - 20] \times 10^{-3} \frac{\mu m}{ms}$, surrounded by an extracellular space with a volume fraction of 0.25[13,14,35]. Therefore, the two exchange mechanisms cannot be easily disentangled using time-dependent SDE measurements. Using advanced diffusion encodings, e.g. double diffusion encoding (DDE) and free gradient waveforms, might be needed to resolve the confounding effects of these mechanisms, as we propose in another study which complements the current one[122].

We further investigated the impact of shaft morphology using beaded and undulated substrates. While the modified Kärger model accurately describes diffusion signals from straight spiny dendritic branches, it fails to capture the signals from beaded and undulated geometries due to additional restriction effects, unaccounted for by the model. This finding is consistent with previous literature[123]. Notably, signals from purely undulated branches align with the model only up to a diffusion time $t_d \leq 150$ ms as shown in **Figure 6.** In these cases, the estimated spine-to-shaft exchange time $\tau_{sp \to sh}$ matches the theoretical values $\tau_{sp \to sh}$. However, a significant increase in the fitted value is observed for highly undulated branches ($N_{und} \geq 1.5$ or $\lambda_{und} \leq 60$ μm ), due to dominating restriction effects along the axial direction.

## 5.2   Impact of dendritic spine morphology and shaft size

Dendritic spine morphology plays a crucial role in diffusion-mediated exchange between intracellular sub-compartments (i.e., from the dendritic spine to the shaft). Based on previous studies investigating dendritic spine morphology[93,96,119,124,125], we estimated the residence times of particles within spines with different morphologies using Eqs.(1) and (3) (see also **Figure 7C**). These equations indicate that while increasing spine neck length and head size prolongs the residence time, increasing neck radius has the opposite effect – as expected as this reduced the surface available for exchange. The interplay of these three factors determines residence time estimations, leading to degeneracy in disentangling individual geometric features. **Figure 7C** illustrates this degeneracy through the $\tau_{sp \to sh}$ isocurves, showing that filopodia-like spines can exhibit the same residence time as mushroom-like spines, and lead to same total exchange times for the system if they have same total volume fraction and SVR. Given the diverse morphologies and classifications of dendritic spines[51–53,126–128], diffusion-exchange analysis alone remains insufficient and ambiguous for precisely characterizing spine morphology. In contrast, our results suggest that the total spine volume fraction may be measurable if diffusion-mediated and permeative exchange mechanisms can be disentangled; but this requires further developments, beyond simple SDE measurements, as discussed in ref.[122].

Regarding the shaft-to-spine exchange, our findings demonstrate that theoretical models of shaft-to-spine exchange, based on the narrow escape problem and Kärger formalism, may systematically overestimate exchange times due to violated assumptions in biologically realistic regimes. The theory of narrow escape problem requires $R_{neck} \ll R_{shaft}$ and negligible radial transit time (the time of particles travelling from the radial distance $R_{shaft} - \varepsilon$ to the pore, with $\varepsilon = \frac{\pi R_{neck}}{4}$)[87,92,94]. For spines, $R_{neck} \lesssim R_{shaft}$ makes ε comparable to $R_{shaft}$, shortening residence times and invalidating the assumption of negligible transit time. Additionally, exponential models (such as the Kärger model) assume constant concentration at exchange points, an approximation that fails when diffusion is hindered (e.g., by long spine necks) or supply pools are small, causing concentration gradients that reduce flux over time, violating the assumptions of the Kärger model. These results highlight critical constraints in applying classical exchange theories to dendritic spine systems and emphasize the need for refined biophysical models that account for finite-size effects and dynamic concentration gradients in restricted geometries.

## 5.3 Characterizing the bias on diffusion exchange estimates

Our simulations using an extended three-compartment Kärger model demonstrate that both diffusion-mediated exchange between spines and shaft and membrane permeability significantly influence diffusion signals in GM. The NEXI model accurately fits these simulated signals (**Figure 9A**), with sensitivity maps revealing the degenerate nature of exchange time estimates ($t_{ex}^{NEXI}$) arising from both mechanisms (isocurves in **Figure 9B**). This degeneracy aligns with known limitations of NEXI/SMEX models in separating restriction from exchange effects[13,14,37,38], cautioning against interpreting exchange times purely as permeative membrane effects.

Quantitative analysis (**Figure 9C-D**) shows that while permeative exchange and spine volume fraction both contribute to observed exchange effects, their individual contributions cannot be disentangled. This is because the increase in total spine volume fraction not only enhances diffusion-mediated exchange but also increases the dendritic surface-to-volume ratio, further amplifying permeative exchange. Nevertheless, our results suggest spine density variations alone can alter $t_{ex}^{NEXI}$ by up to 80%. These results can help interpret variations in exchange time measured in conditions or pathologies that are known to alter spine density. For example, in autism spectrum disorder, where spine density in temporal lobe deep layers is ~20% higher than in controls ($\rho \sim 1\ \mu m^{-1}$ *versus* $\rho \sim 0.8\ \mu m^{-1}$; corresponding to $\nu \sim 0.17$ and $\nu \sim 0.14$ respectively[75]), our results suggest a corresponding ~10% decrease in $t_{ex}^{NEXI}$; potentially misattributed to increased permeability.

The estimated exchange times using NEXI for reasonable total spine volume fraction 10-20% match experimental estimates in human cortex, i.e. 10-150 ms [13–15,36]. Regional differences in spine density may further explain variations in reported exchange times. The prefrontal cortex, with higher spine densities (~3 $\mu m^{-1}$), likely exhibits shorter exchange times than the primary visual cortex (~0.5 $\mu m^{-1}$)[73,129], consistent with observed longer exchange times (~100–150 ms) in the latter[35]. Similarly, the hippocampus and cerebellar cortex - regions with high spine densities[130] - show shorter exchange times (~50–100 ms)[14,15,35], while the spiny Purkinje cells of the cerebellum may contribute to its elevated exchange rates[23].

Collectively, these findings underscore dendritic spines as a critical, though often overlooked, source of non-permeative exchange. Alongside myelination and other microstructural factors[16], spine density variations likely drive the observed regional differences in exchange estimates across the brain[13,14,23,36].

## 5.4 Perpendicular signal component in dendritic spines

In general, the perpendicular signal component of diffusing water molecules in the GM is considered negligible due to small radii of neurites[13,14,32,118,123,131] and motional averaging of the diffusion process in the clinical measurements[14,38,132,133]. Consequently, neurites in GM are often assumed to exhibit non-decaying perpendicular diffusion signals and are modeled as randomly oriented sticks. However, studies employing ultra-high diffusion-weighting have not provided strong evidence supporting the validity of the stick model in GM[118,131,134]. On the contrary, the presence of dendritic spines may violate this assumption by increasing the effective fiber radius at higher spine densities[120], thereby introducing additional signal bias and $q - t$ dependence in the direction perpendicular to the main axis of the shaft. In this study, while our primary investigation focuses on the diffusion and exchange dynamics of dendritic spines, we also wish to highlight an important observation from our simulations: the non-negligible signal decay of the perpendicular signal component. We believe this finding is of interest to the diffusion community. Examining the perpendicular signal component of the Monte Carlo simulations, we found that there is considerable signal attenuation up to 40% for typical spine densities (0.5 – 3 $\mu m^{-1}$) at short and moderate diffusion times. The **Supplementary Figure S5** shows the perpendicular signal attenuations ($S_\perp$) for several spine densities from the Monte Carlo simulations with narrow (A) and wide (B) pulses. The signal decay in the motional averaging regime[132,133] varies between 10% to 20% for typical spine densities (see **Supplementary Figure S5B**), indicating that $S_\perp$ has non-negligible impact in the diffusion analysis. The solid lines in **Supplementary Figure S5A** and **B** are the fitted lines to the cylinder model[135]. The estimated effective fiber radii in the spiny dendritic branches are also illustrated as a function of spine density for narrow (solid lines) and wide (dotted lines) pulse simulation settings in the **Supplementary Figure S5C**. The dashed line presents the maximum available distance along the spines ($L_{neck} + 2R_{head} + R_{shaft}$ = 2.6 $\mu$m). Estimated effective fiber radii show sensitivity to different spine densities, increasing with increasing spine density until converging to a limit value that matches the expected one for narrow pulse case. The larger effective dendritic branch size due to the presence of spines agrees with previous work studying the spines and leaflets[120]. Palombo et al. reported larger branch sizes estimated due to the total spine length and density. Therefore, the potential signal attenuation in the direction perpendicular to the neurites might be significant in dMRI measurements due to the dendritic spines in the GM. To address this potential bias in the context of exchange estimation, experimental designs that vary the exchange encoding while keeping the restricting encoding constant would be advised[24].

## 5.5 Limitations

We acknowledge a few limitations of the current study. Our digital models do not include various spine populations in addition to mushroom-like, which are instead typically present in the real branches, such as immature filopodia or spike-like spines. Nonetheless, the synthetic dendrites feature the same exchange dynamics as the real ones. In addition, our simulations were performed in spiny dendritic branches with varying spine density, but do not accurately simulate the macroscopic voxel. Hence, our findings resulted from oversimplified simulations. For instance, while previous studies[7,13] have reported negligible or distinguishable effects of cell bodies on exchange dynamics between the cell body and neurites, our analysis does not account for the restriction effects of cell bodies on the diffusion process. More realistic diffusion simulations might be achieved by employing numerical gray matter phantoms[136] after the inclusion of dendritic spines.

Our diffusion simulations were designed with simple diffusion encodings[34] and studied short and wide pulse acquisitions. Although our results align with the literature regarding the overestimation observed in exchange estimates with wide pulses[14,35], the degeneracy in estimating diffusion metrics of restriction and exchange remained unresolved in our SDE based simulations[13,14,37,38,137]. More advanced acquisition techniques address this issue by incorporating additional experimental dimensions[23,24,26,38,48], which may help resolve the degeneracy between total spine volume fraction and membrane permeability observed in estimated exchange times in our analysis, and it is the focus of another study[122] which complements the present one.

Finally, the proposed extended Kärger model of three compartments neglects the diffusion signal decay in the dendritic spines and considers it as dot compartment for simplification. This assumption in our proposed model might overestimate the restriction effect on the molecular diffusion process. Further corroborations are required to assess the measurable diffusion in dendritic spines and its potential impact on the interplay between restriction and exchange.

## 6   Conclusion

Our study demonstrates that the time-dependent SDE signal arising from diffusion within impermeable spiny dendrites is indistinguishable from that resulting from permeative exchange. We show that a modified two-compartment Kärger model accurately captures the time-dependent SDE signal along the dendritic shaft in spiny dendritic branches but yields biased estimates of exchange times. Notably, exchange estimates within spiny dendritic branches cannot be uniquely attributed to specific spine morphologies due to degeneracy but instead reflect the overall spine density per unit volume. We proposed an extended three-compartment Kärger model, which incorporates the contribution of dendritic spines, to simulate exchange between dendritic shaft, dendritic spines and extracellular space. Our analysis using the extended Kärger model revealed that diffusion-mediated exchange within spiny dendrites can introduce up to 80% bias in NEXI/SMEX estimates of exchange time. Therefore, the presence of spiny dendrites and unaccounted diffusion-mediated exchange can bias NEXI/SMEX estimates of permeative exchange times in gray matter, depending on the total spine volume fraction. Although useful to study the interweaved impact of both diffusion-mediated and permeative exchange, the proposed three-compartment Kärger model cannot disentangle the respective contributions of membrane permeability and total spine volume fraction. Our findings highlight the need for careful interpretation of diffusion MRI-based exchange estimates, particularly when attributing them solely to membrane permeability; and highlight the unmet need for better methods and modelling approaches to disentangle permeative and diffusion-mediated exchange mechanisms.

## Acknowledgements

This work, KS and MP are supported by UKRI Future Leaders Fellowship (MR/T020296/2). AC and MN are supported by Swedish Research Council (2024-04968) and Hjärnfonden (FO2024-0335-HK-73).

## CRediT authorship contribution statement

Kadir Şimşek: Conceptualization, Formal analysis, Methodology, Investigation, Writing - original draft;

Arthur Chakwizira: Writing - reviewing draft

Markus Nilsson: Writing - reviewing draft

Marco Palombo; Conceptualization, Methodology, Funding, Writing - original draft, Supervision

## Conflict of Interest Statement

The Authors have no conflict of interest to declare.

## Data Availability Statement

All the data and analysis codes underpinning the results presented here can be found upon publication in the Cardiff University data catalogue and on Github: https://github.com/kdrsimsek.

# Supplementary Figures & Tables

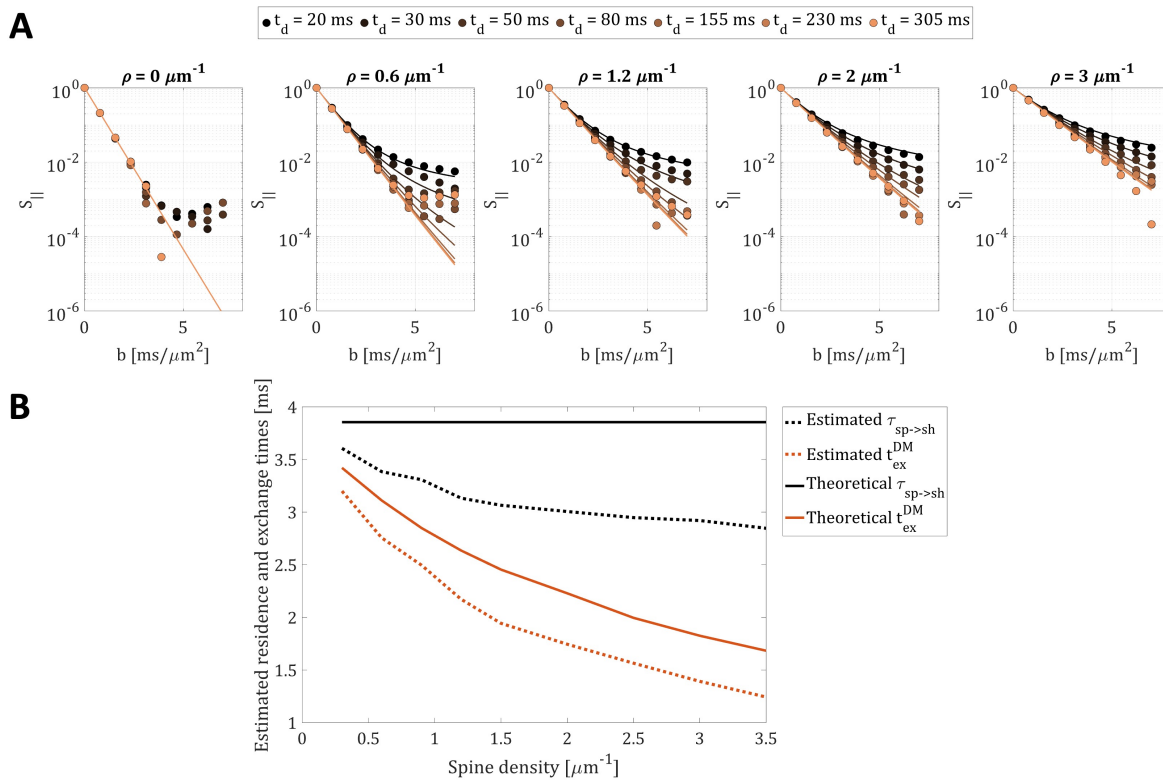

**Figure S1**: *Theoretical signal predictions from the modified Kärger model are overlaid on simulated parallel diffusion signals (S$_{\parallel}$) across all diffusion times using the wide pulse scheme (A). Additionally, model predictions are shown over a range of spine densities (ρ), illustrating the impact of ρ on diffusion behavior. (B) The panel shows the estimated residence time $\tau_{sp \to sh}$ (black) and exchange time $t_{ex}$ (orange) obtained using the modified Kärger model, under the wide gradient pulse settings. These estimates are compared with the corresponding theoretical predictions ($\tau_{sp \to sh}$ and $t_{ex}^{DM}$). The dotted lines present the estimations of the modified Kärger model while the theoretical predictions are displayed as solid lines for both spine-to-shaft residence times and exchange times. Both $\tau_{sp \to sh}$ and $t_{ex}$ estimates from the model are consistently underestimated compared to the theoretical predictions under the wide pulse scheme.*

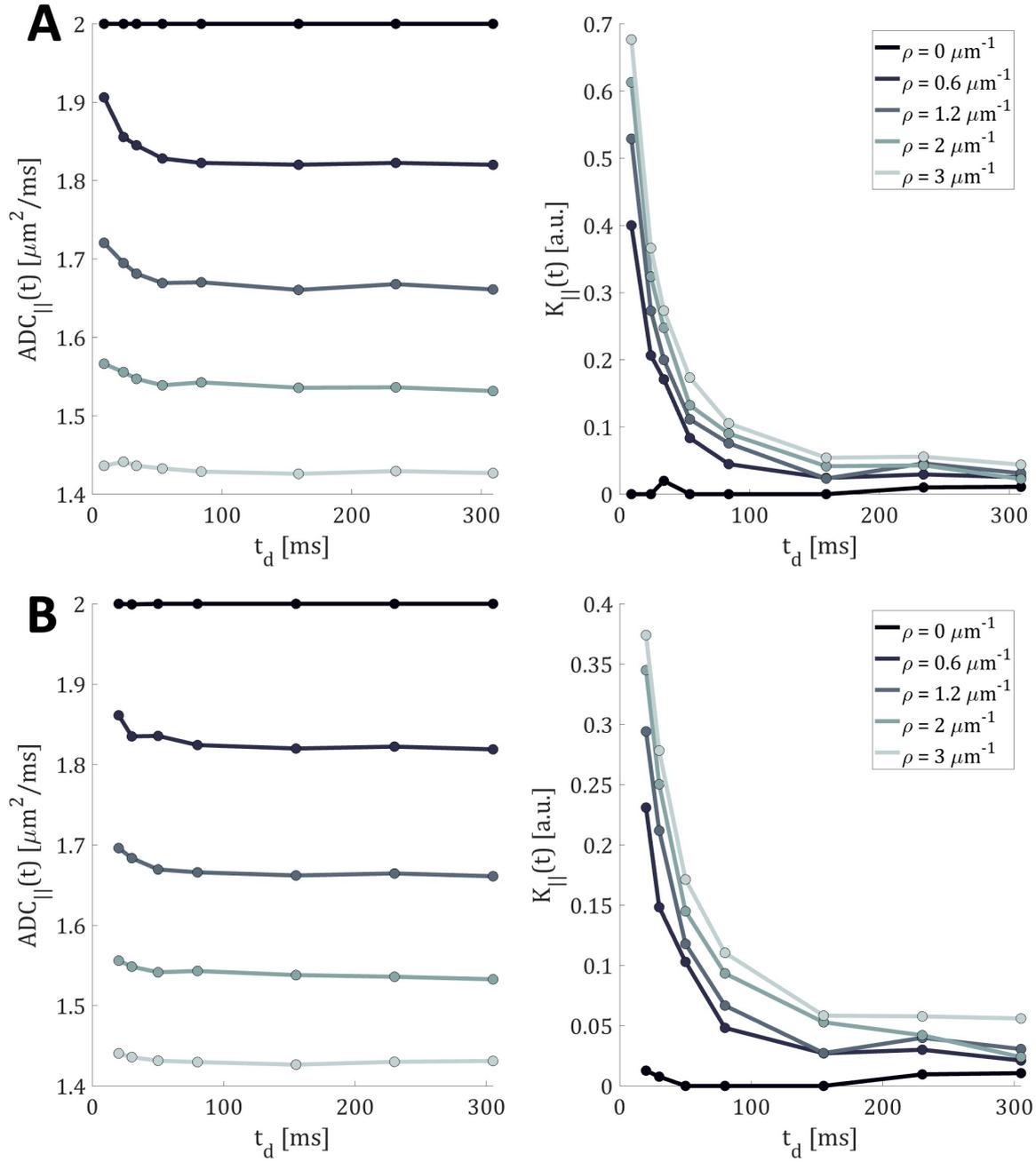

**Figure S2**: *Time dependence of the apparent diffusion coefficient (ADC) and kurtosis (K) values estimated from simulated parallel diffusion signals ($S_\parallel$) along the dendritic shaft for various spine densities, using narrow (A) and wide (B) gradient pulse schemes. The kurtosis K decreases monotonically with time for spine densities $\rho > 0\ \mu m^{-1}$ which indicates that the barrier-limited exchange condition is satisfied up to $t_d \leq 154\ ms$.*

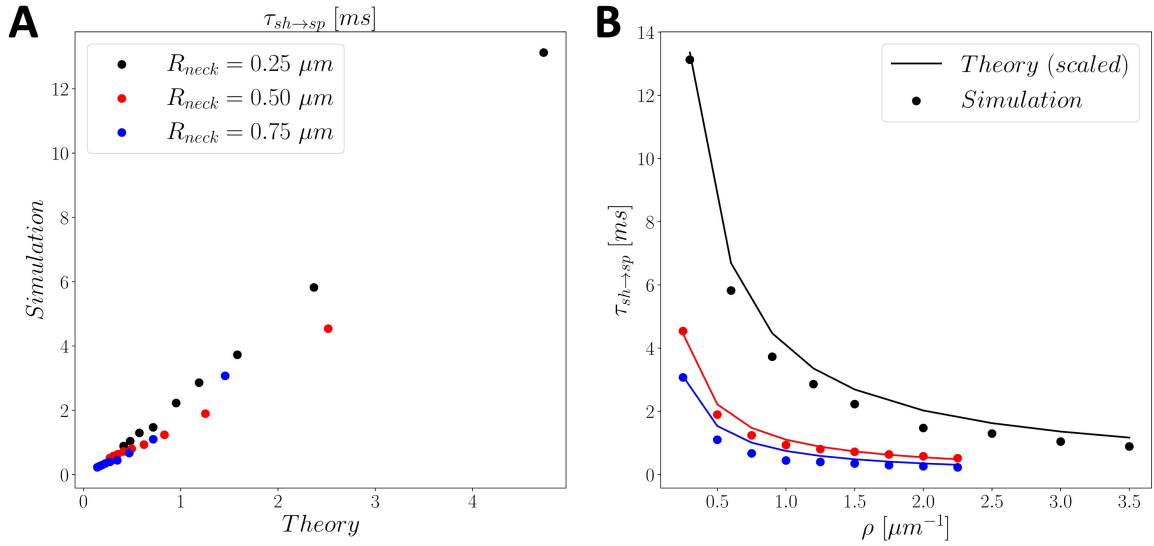

**Figure S3**: *(A) Estimated residence times from shaft-to-spine ($\tau_{sh \to sp}$) for different spine densities are compared with the theoretical results (Eq.(3)). The color coding indicates substrate sets with different pore sizes (e.g. spine neck radius $R_{neck}$). The theoretical residence times are a few orders of magnitude smaller than those obtained from the simulation results. (B) Illustration of residence times of substrate sets as a function of time. Here, the theoretical values are scaled with the slope estimated in (A). Despite a few orders of magnitude difference in residence times obtained from the simulations, the characteristic decay as a function of spine density agrees with the theory.*

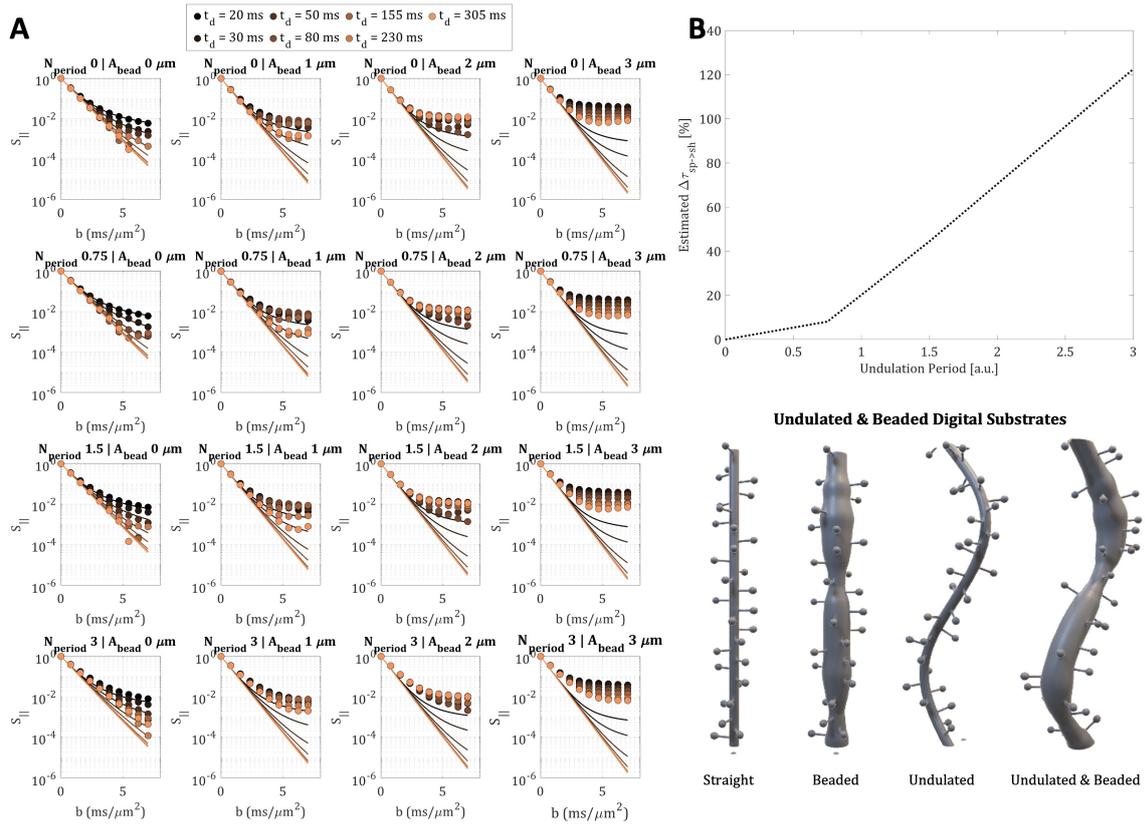

**Figure S4**: *The diffusion signals in the parallel direction obtained from undulated and beaded synthetic substrates with spine density ρ = 1 μm⁻¹ are shown in the figure. The diffusion signals generated using wide pulse gradient scheme are presented in (A). At long diffusion times ($t_d > 150\ ms$), the restriction effect starts dominating the diffusion signal. In (B), the percentage change in the spine-to-shaft residence times ($\Delta\tau_{sp\rightarrow sh}$), estimated by fitting the modified Kärger model, is shown as a function of undulation period $N_{und}$, relative to a reference branch without undulation or beading. For undulation per $N_{und} < 1.5$, the $\tau_{sp\rightarrow sh}$ results present less than 10% impact of undulation on the estimated residence times.*

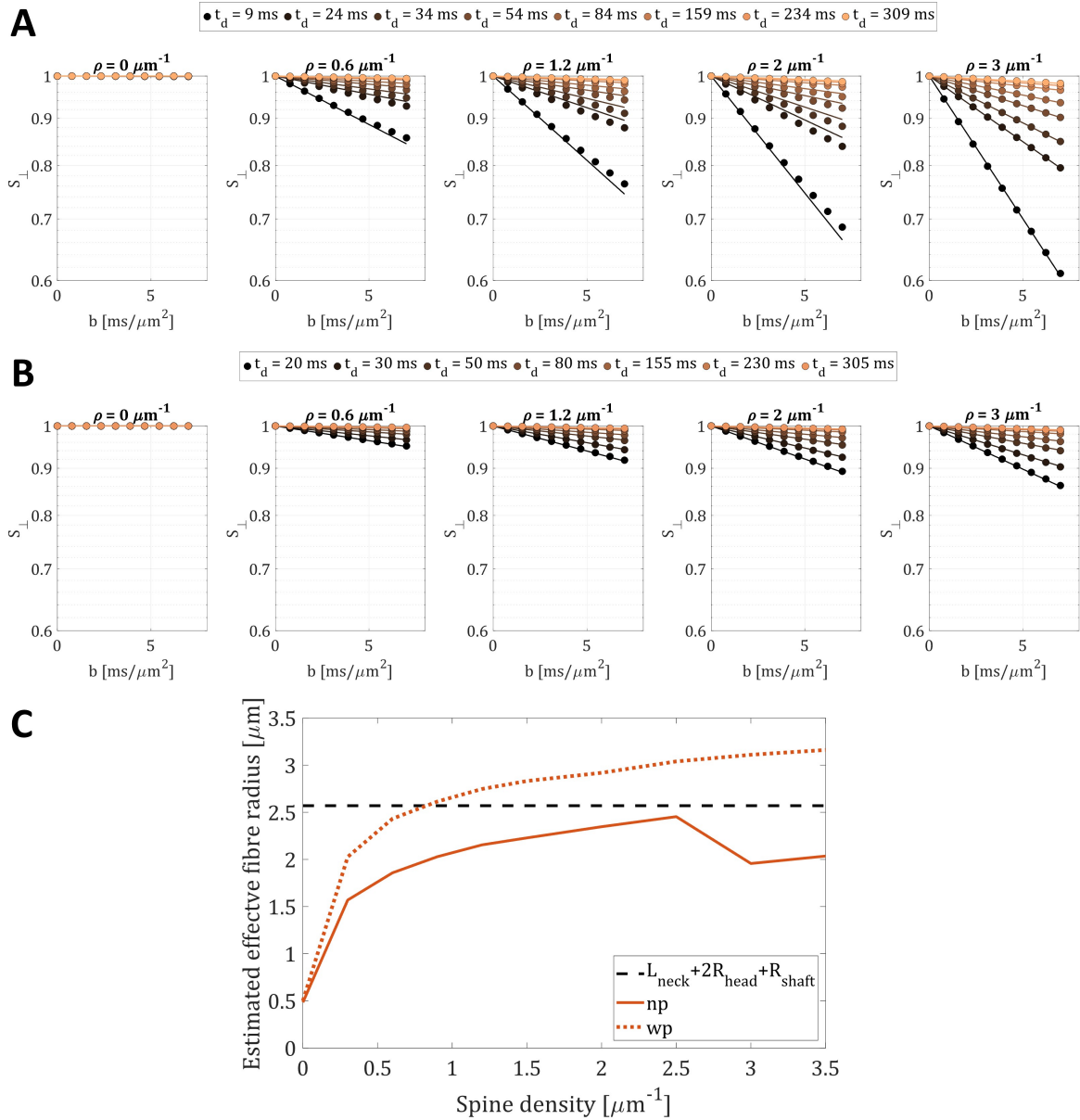

**Figure S5**: *The perpendicular signal component ($S_\perp$) of the simulated diffusion signals for several spine densities are shown in the figure for narrow (A) and wide (B) gradient pulses. For typical spine densities, the signal attenuation $S_\perp$ is not negligible and can be up to 30%, especially at short and moderate diffusion times. The solid lines are obtained from fitting $S_\perp$ to randomly oriented cylinders model[135]. (C) Estimated effective cylinder radii (solid: narrow pulse and dotted: wide pulse) are shown as a function of spine density. Dashed lines indicate the maximum available distance ($L_{neck} + 2R_{head} + R_{shaft}$) in the spiny dendritic branches. Both effective radii depend on the spine density and have an increasing trend with increasing spine density up to 2 spines per micron.*

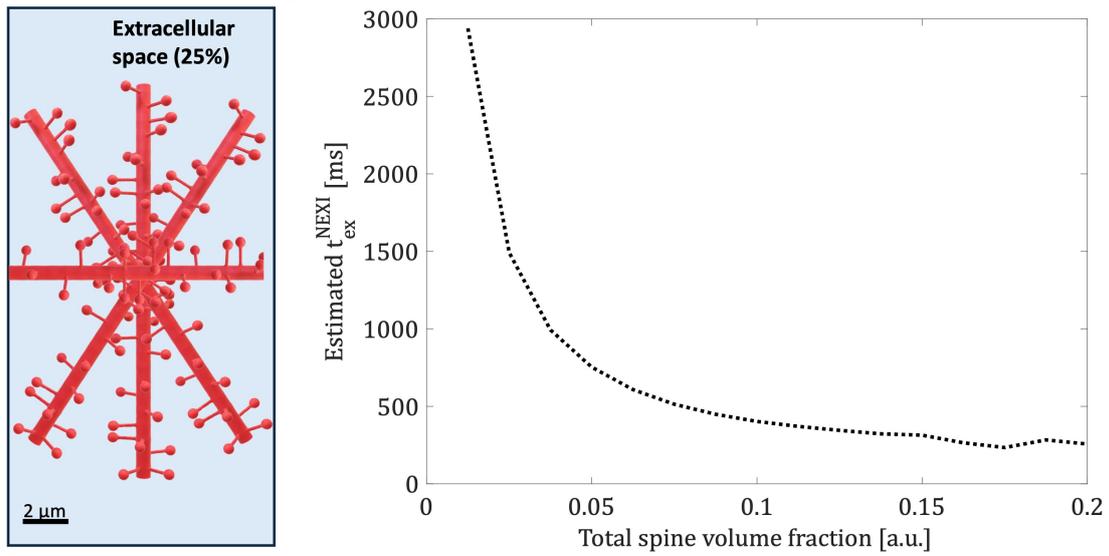

***Figure S6:*** *Illustration of diffusion-mediated exchange within spiny branches. The figure shows the bias in NEXI estimates of $t_{ex}^{NEXI}$ when spines and exchange with spines is not considered, for wide gradient schemes. We generated the signal using 75% spiny dendrite + 25% isotropic gaussian extracellular space with diffusivity $1.2 \frac{\mu m^2}{ms}$ and fitted the NEXI model to it for different spine densities. The data points are smoothed with a one-point window for better visualization.*